\documentclass[12pt,onecolumn,draft]{IEEEtran}
\usepackage[margin=1in]{geometry}
\usepackage[final]{graphicx}
\usepackage{amsmath,amssymb,amsfonts,latexsym,cite}

\normalsize \tolerance 5000

\def\figref#1{Fig.~\ref{#1}}
\def\secref#1{Section~\ref{#1}}

\def\lemref#1{Lemma~\ref{#1}}
\def\thmref#1{Theorem~\ref{#1}}
\def\qed{\rule{.5ex}{.5em}}

\def\iid{i.i.d.\ }

\def\bdA{\boldsymbol{A}}
\def\bdB{\boldsymbol{B}}
\def\bdC{\boldsymbol{C}}
\def\bdD{\boldsymbol{D}}
\def\bdE{\boldsymbol{E}}

\def\bdG{\boldsymbol{G}}
\def\bdH{\boldsymbol{H}}
\def\bdI{\boldsymbol{I}}

\def\bdP{\boldsymbol{P}}
\def\bdQ{\boldsymbol{Q}}
\def\bdR{\boldsymbol{R}}

\def\bdV{\boldsymbol{V}}
\def\bdW{\boldsymbol{W}}

\def\bde{\boldsymbol{e}}

\def\bdg{\boldsymbol{g}}
\def\bdh{\boldsymbol{h}}

\def\bdn{\boldsymbol{n}}

\def\bdr{\boldsymbol{r}}

\def\bdv{\boldsymbol{v}}

\def\bdx{\boldsymbol{x}}
\def\bdy{\boldsymbol{y}}
\def\bdz{\boldsymbol{z}}

\def\mD{\mathcal{D}}

\def\mH{\mathcal{H}}
\def\mI{\mathcal{I}}

\def\bdmV{\boldsymbol{\mathcal{V}}}

\def\bdzeros{\boldsymbol{0}}
\def\bdones{\boldsymbol{1}}

\def\bdLambda{\boldsymbol{\Lambda}}

\def\CN#1{\mathcal{CN}(#1)} 

\def\va{\bdy_a}
\def\vb{\bdy_b}
\def\vc{\bdy_c}
\def\vbc{\bdy_{bc}}
\def\spF{\textsf{F}}
\def\spS{\textsf{S}}

\def\dg{\dagger}
\def\diag{\text {diag}}

\def\E{\text{E}}
\def\rank{\text{rank}}
\def\vt{\text{vec}}
\def\vand{\bdmV}

\def\bydef{:=}
\def\N#1{N_{#1}}
\def\M#1{M_{#1}}

\def\R#1{R_{#1}}

\long\def\comment#1{}
\newtheorem{theorem}{Theorem}
\newtheorem{lemma}{Lemma}

\def\logP{\log(P)}
\def\ologP{o(\log(P))}
\def\Hall{\underline{\bdH}_{11}}

\def\tdWall{\underline{\tilde\bdW}_{11}}

\def\tdVall{\underline{\tilde\bdV}_{11}}

\def\tdLmdall{\underline{\tilde\bdLambda}_{11}}

\def\lb{\ifCLASSOPTIONonecolumn\else\\ \fi}

\begin{document}

\title{Degrees of Freedom Regions of Two-User MIMO Z and Full
Interference Channels: The Benefit of Reconfigurable Antennas}

\author{Lei~Ke, \IEEEmembership{Student Member,~IEEE}, and \\
Zhengdao~Wang (Contact Author), \IEEEmembership{Senior Member,~IEEE}
\thanks{L. Ke and Z. Wang are with the Department of
Electrical and Computer Engineering, Iowa State University, Ames, IA 50011 USA
(e-mail: kelei@iastate.edu; zhengdao@iastate.edu).}
\thanks{Part of the work in this paper was accepted by IEEE GLOBECOM 2010.}
}

%

\maketitle

\begin{abstract}
We study the degrees of freedom (DoF) regions of two-user multiple-input
multiple-output (MIMO) Z and full interference channels in this paper. We
assume that the receivers always have perfect channel state information. We
first derive the DoF region of Z interference channel with channel state
information at transmitter (CSIT). For full interference channel without CSIT,
the DoF region has been fully characterized recently and it is shown that the
previously known outer bound is not achievable. In this work, we investigate
the no-CSIT case further by assuming that the transmitter has the ability of
antenna mode switching. We obtain the DoF region as a function of the number
of available antenna modes and reveal the incremental gain in DoF that each
extra antenna mode can bring. It is shown that in certain cases the
reconfigurable antennas can bring extra DoF gains. In these cases, the DoF
region is maximized when the number of modes is at least equal to the
number of receive antennas at the corresponding receiver, in which case the
previously outer bound is achieved. In all cases, we propose systematic
constructions of the beamforming and nulling matrices for achieving the DoF
region. The constructions bear an interesting space-frequency interpretation.

\end{abstract}

\newpage

\section{Introduction} Characterizing the capacity region of interference
channel has been a long open problem. Many researchers investigated this
important problem, and the capacity regions of certain interference channels
are known when the interference is strong, e.g. \cite{carl75,sato81,gaco82}.
However, when the interference is not strong, the capacity region is still
unknown. Recent progress reveals the capacity region for two-user interference
channel to within one bit \cite{ettw08}, and after that the sum capacity for
very weak interference channel is settled \cite{shkc08c,srve08c,mokh08c}.
Recently, a deterministic channel model has been proposed and used to explore
the capacity of Gaussian interference network \cite{gbdt08,gbpt08,sadt09} such
that the gap to capacity region can be bounded up to a constant value.

When it comes to multiple-input multiple-output (MIMO) networks, the capacity
regions of certain MIMO interference channels are known \cite{avvv09,sckp09}.
Instead of trying to characterize the capacity region completely, the degrees
of freedom (DoF) region characterizes how capacity scales with transmit power
as the signal-to-noise ratio goes to infinity.

It is well-known that in certain cases, the absence of channel state
information at transmitter (CSIT) will not affect the DoF for MIMO networks,
e.g., in the multiple access channel \cite{gjjv03}. In other cases, CSIT does
play an important role. For example, using interference alignment scheme, it
is shown that the total DoF of a $K$-user MIMO interference channel is $MK/2$,
where $M$ is the number of antennas of each user \cite{caja08}. The key idea
is to pack interferences from multiple sources so as to reduce the
dimensionality of signal space spanned by interference.

The DoF region of two-user MIMO interference channel with CSIT has been
obtained in \cite{jafa07}, where it is shown that zero forcing is enough to
achieve the DoF region. However, it is a different story in two-user MIMO X
channel, where each transmitter has a message to every receiver. In
\cite{jash08} it is shown that interference alignment is the key to achieving
the DoF region of MIMO X network. The DoF region of two-user MIMO broadcast
channel and interference channel without CSIT are considered in \cite{hjsv09},
where there is an uneven trade-off between the two users. Except for a special
case, the DoF region for the interference channel is known and achievable.
Similar, but more general result of isotropic fading channel can be found in
\cite{zhgu09c}. The DoF regions of the $K$-user MIMO broadcast, interference
and cognitive radio channels are derived in \cite{cvmv09} for some cases.
However, the special case in \cite{hjsv09} remains unsolved.

When only one of the two transmitter-receiver pairs is subject to
interference, the interference channel is termed as \emph{Z interference
channel} (ZIC). To avoid confusion, we will call the channel where both pairs
are subject to interference the \emph{full interference channel} (FIC). The
capacity region of MIMO Gaussian ZIC is established in \cite{sckp09c} under
very strong interference and aligned strong interference assumptions. In
\cite{yzdg09}, the authors considered the capacity region of a single antenna
ZIC without CSIT using deterministic approach.

Recently, it is shown in \cite{jafar09} that if the channel is staggered block
fading, we can explore the channel correlation structure to do interference
alignment, where the upper bound in the converse can be achieved in some
special cases. For example, it is shown that for two-user MIMO staggered block
fading FIC with 1 and 3 antennas at transmitters, 2 and 4 antennas at their
corresponding receivers and without CSIT, the DoF pair $(1, 1.5)$ can be
achieved. The idea was further clarified in \cite{cwtgs10}, where a blind
interference alignment scheme is also proposed for $K$-user multiple-input
single-output (MISO) broadcast channel to achieve DoF outer bound when CSIT is
absent. Also recently, it is shown in \cite{yzdg10a} that the previous outer
bound is not tight when the channels are independent and identical distributed
(i.i.d.) over time and isotropic over spatial domain. So by now the DoF region
of two-user MIMO FIC is completely known for both the case with CSIT and the
no CSIT case (receiver-side CSI, or CSIR, is always assumed available),
provided that the channel is \iid over time and isotropic over spatial domain.
However, when the channel is not \iid over time such as in the ``staggered''
fading channels \cite{jafar09}, the DoF could be larger.

In this paper, we consider the ZIC channel with CSIT, and both ZIC and FIC
without CSIT but with reconfigurable antennas. Specifically, we obtain the DoF
regions for the cases of:
\begin{enumerate}
\item ZIC with CSIT. We show that zero forcing is sufficient for achieving the
DoF region in this case (\thmref{thm.zic.csit}).

\item ZIC and FIC when transmitter one has the number $K$ of antennas modes at
least equal to $N_1$ (Theorems~\ref{thm.zic.nocsit} and \ref{thm.fic.nocsit}).
Increasing $K$ beyond $N_1$ does not bring more gains in DoF.

\item ZIC and FIC when $M_1\le K<N_1$, in which case each additional antenna
mode brings an incremental gain on the DoF region (\thmref{thm.lessmodes}).

\end{enumerate}
We present joint beamforming and nulling schemes to achieve the DoF region in
all cases. When reconfigurable antennas are used, our proposed schemes have an
interesting space-frequency coding explanation.

The rest of the paper is organized as follows. We first present the system
model in \secref{sec.model}. Known results on the DoF region of two-user MIMO
FIC are also briefly reviewed. The DoF region of ZIC with CSIT is discussed in
\secref{sec.zic.csit}. The DoF regions of ZIC and FIC without CSIT when there
are enough antenna modes are investigated in \secref{sec.kgeqn1}. When there
are not enough modes, the DoF region is given in \secref{sec.lessmodes}.
Finally, \secref{sec.conc} concludes this paper.

Notation: boldface uppercase (lowercase) letters denote matrices (vectors).
$\mathbb{R},\mathbb{Z}, \mathbb{C}$ are the real, integer and complex numbers
sets. $\CN{0,1}$ denotes a circularly symmetric complex Gaussian (CSCG)
distribution with zero mean and unit variance. We use $\bdA\otimes \bdB$ to
denote the Kronecker product between $\bdA$ and $\bdB$. $\bdzeros$ and
$\bdones$ denote all one and all zero matrices (vectors), respectively.
$\bdA^T$ and $\bdA^\dg$ denote the transpose and Hermitian of $\bdA$,
respectively. We also use notation like $\bdA_{m\times n}$ to emphasize that
$\bdA$ is of size $m \times n$. We use $\bdI_m$ to denote a size $m\times m$
identity matrix and $\bdones_m$ to denote an all-one column vector with length
$m$. Denote $\bdg_n(a)\bydef[1, a,a^2,\dots,a^{n-1}]^{T}$. A size $n\times m$
Vandermonde matrix based on a set of element $\{a_1,a_2,\dots,a_m\}$ is
defined as
$\vand_n(a_1,a_2,\dots,a_m)=[\bdg_n(a_1),\bdg_n(a_2),\dots,\bdg_n(a_m)]$. We
use $\mI(\bdx;\bdy)$ to denote the mutual information between $\bdx$ and
$\bdy$. The differential entropy of a continuous random variable $\bdx$ is
denoted as $\mH(\bdx)$.

\section{System Model and Known Results}\label{sec.model}

\subsection{Channel Model}

Consider a MIMO interference channel with two transmitters and two receivers,
the number of transmit (receive) antennas at the $i$th transmitter (receiver)
is denoted as $M_i$ ($N_i$), $i\in \{1,2\}$. The system is termed as an
($\M1,\N1,\M2,\N2$) system, which can be described as
\begin{align}
\bdy_1(t)=\bdH_{11}(t)\bdx_1(t)+\bdH_{12}(t)\bdx_2(t)+\bdz_1(t)\\
\bdy_2(t)=\bdH_{21}(t)\bdx_1(t)+\bdH_{22}(t)\bdx_2(t)+\bdz_2(t)
\end{align}
where $t$ is the time index, $\bdy_i(t)\in \mathbb{C}^{N_i}$, $\bdz_i(t)\in
\mathbb{C}^{N_i}$ are the received signal and additive noise of receiver $i$,
respectively. The entries of $\bdz_i(t)$ are independent and identically
$\CN{0,1}$ distributed in both time and space. The channel between the $i$th
transmitter and the $j$th receiver is denoted as $\bdH_{ji}(t)\in
\mathbb{C}^{N_j \times M_i}$. We assume the probability of $\bdH_{ji}(t)$
belonging to any subset of $\mathbb{C}^{N_j \times M_i}$ that has zero
Lebesgue measure is zero. For the two-user MIMO ZIC, $\bdH_{21}(t)=0$.
$\bdx_i(t) \in \mathbb{C}^{M_i}$ is the input signal at transmitter $i$ and
$\bdx_1(t)$ is independent of $\bdx_2(t) $. The transmitted signals satisfy
the following power constraint:
\begin{align}
\E(||\bdx_i(t)||^2)\leq P \quad i=1,2.
\end{align}

Denote the capacity region of the two-user MIMO system as $C(P)$, which
contains all the rate pairs $(R_1,R_2)$ such that the corresponding
probability of error can approach zero as coding length increases. The DoF
region is defined as follows \cite{hjsv09}
\begin{align*}
\mD \bydef \left\{(d_1,d_2)\in \mathbb R^+_2: \exists
(\R1(P),\R2(P))\in C(P),\right. \lb
\left. \text{such that } d_i
=\lim_{P\to\infty}\frac{R_i(P)}{\log(P)}  , \quad i=1,2 \right \}.
\end{align*}

\subsection{Reconfigurable antennas}

Assume the CSI at receiver (CSIR) is always available. We would like to study
the DoF regions of MIMO FIC and ZIC with or without CSIT under an additional
assumption that one transmitter is equipped with reconfigurable antennas. The
reconfigurable antennas are different from the conventional antennas as they
can be switched to different pre-determined modes so that the channel
fluctuation can be introduced artificially. Similar to \cite{cwtgs10}, we use
reconfigurable antennas to explore multiplexing gain other than diversity
gain.

We assume that only one transmitter is equipped with reconfigurable antennas.
We define one antenna \emph{mode} as one possible configuration of a single
transmit antenna such that by switching to a different mode, the channel
between this transmit antenna and all receive antennas is changed. Different
antenna modes can be realized via spatially separated physical antennas, or
the same physical antenna excited with different polarizations, and so on. The
benefit of antenna mode switching lies in the fact that channel variation can
be artificially created, without the need to increase the number of RF chains.
We let $K$ denote the total number of antenna modes available at the
transmitter with reconfigurable antennas (usually transmitter one).

We make the following assumption of the channel in this paper: the channel is
block fading with coherent length of $L$ symbols. Within each coherent block,
the channels between all the transmitter modes and the receive antennas remain
constant. The channels between the $K$ modes of the reconfigurable transmitter
and both receivers are isotropic, in the sense of \cite{yzdg10a}. From block
to block, the channel changes independently.

When $K>M$, the transmitter has the freedom to use different modes at
different slots. For a given antenna mode usage pattern over the length of a
whole coherent block, the effective channel for the whole block is not
isotropic fading and not \iid over the time slots within the block.

One may view our model approximately as a transition from an effective channel
where all the links have exactly the same coherent time as in \cite{yzdg10a}
to an effective channel where the links do not have the same coherent time
\cite{jafar09}. However, there are two important distinctions between antenna
mode switching and variation of channel coherence time: i) Antenna switching
can be initiated at will at the transmitter, whereas channel coherence
structure is in general not controllable. ii) The resulting equivalent channel
from antenna mode switching is not ``staggered'' \cite{jafar09}, so methods
therein do not apply here.

\subsection{Known Results on FIC}
\label{sec.ic}

We first present some known results on DoF region of MIMO full interference
channel which will be useful for developing our results.

The total degrees of freedom of two-user MIMO full interference channel with
CSIT is developed in \cite[Theorem 2]{jafa07}, which leads to the following
DoF regions:
\begin{align}
d_i &\leq \min(M_i,N_i), \quad i=1,2; \\
d_1+d_2&\leq \min(\max(N_1,M_2),\max(M_1,N_2),
\N1+\N2,\M1+\M2).
\end{align}

An outer bound of degrees of freedom region of two-user MIMO full interference
channel without CSIT is as follows \cite[Theorem 1]{zhgu09c}:
\begin{align}
&d_i \leq \min(M_i,N_i), \quad i=1,2;  \label{eq.nocsit.upper0}\\
&d_1 \!+\!\frac{\min(\N1,\N2,\M2)}{\min(\N2,\M2)}d_2  \leq
\min(M_1+M_2,N_1);
        \label{eq.nocsit.upper.m}
\\
&\frac{\min(\N1,\N2,\M1)}{\min(\N1,\M1)}d_1\!+\!d_2  \leq
\min(M_1+M_2,N_2).
        \label{eq.nocsit.upper1}
\end{align}
Note that the same result is also given in \cite{hjsv09}, though in a less
compact form.

It is shown in \cite{hjsv09} that the outer bound given in
\eqref{eq.nocsit.upper0}--\eqref{eq.nocsit.upper1} can be achieved by zero
forcing or time sharing except for the case $\M1 < \N1 < \min(\M2,\N2)$, for
which it was not known how to achieve
\begin{align}(d_1,
d_2)=\left(\M1,\frac{\min(\M2,\N2)(\N1-\M1)}{\N1}\right) \label{eq.DoFunknown}
\end{align}
in general. {The cases when $\N1>\N2$ can be converted by switching the user
indices}. It is shown in \cite{yzdg10a} that when the channel is isotropic
fading and i.i.d. over time, the outer bound given in
\eqref{eq.nocsit.upper0}--\eqref{eq.nocsit.upper1} is not tight: if $N_1\leq
N_2$, the DoF region of FIC without CSIT
 can be given as follows:
\begin{align}
&d_i \leq \min(M_i,N_i), \quad i=1,2; \label{eq.nocsit.iso0} \\
&d_1 \!+\!\frac{\min(\N1,\M2)-\alpha}{\min(\N2,\M2)-\alpha}(d_2-\alpha)  \leq
\min(M_1,N_1). \label{eq.nocsit.iso1}
\end{align}
where $\alpha=\min(M_1+M_2,N_1)-\min(M_1,N_1)$. In other words,
\eqref{eq.DoFunknown} is not achievable when $\M1 < \N1 < \min(\M2,\N2)$, as
\eqref{eq.nocsit.iso1} is reduced to
\begin{align} \label{eq.dofzic.simple}
d_1+\frac{M_1}{\min(M_2,N_2)-(N_1-M_1)}d_2&\leq M_1
	+\frac{M_1(N_1-M_1)  }{\min(M_2,N_2)-(N_1-M_1)}
\end{align}
and the DoF pair $(d_1,d_2)=(M_1,N_1-M_1)$ is on the boundary of the DoF
region.

\section{Two-User MIMO ZIC with CSIT} \label{sec.zic.csit}

In this section, we prove the following theorem.

\begin{theorem} [ZIC with CSIT] \label{thm.zic.csit}
\label{thm.z.csit.DoF}The DoF region of a two-user MIMO
Z interference channel with CSIT is described by
\begin{align}
d_i &\leq \min(M_i,N_i), \quad i=1,2;\label{eq.zic.region1}\\
d_1+d_2&\leq \min(\max(N_1,M_2),\N1+\N2,\M1+\M2). \label{eq.zic.region2}
\end{align}\hfill
\end{theorem}

\begin{IEEEproof}
We split the proof into the achievability and converse parts, as the following
two lemmas. The theorem can be proved by showing the regions given by
\lemref{lem.zic.inner} and \lemref{lem.zic.outer} are the same for all the
cases. \hfill
\end{IEEEproof}

\begin{lemma}[Achievability part of \thmref{thm.zic.csit}]
\label{lem.zic.inner}
The following region of two-user MIMO ZIC with CSIT is
achievable:\begin{align} d_i &\leq \min(M_i,N_i), \quad i=1,2; \\ d_1+d_2&\leq
\min(\N1,\M1+\min(\N2,\M2))1({\M2< \N1})
\nonumber\\&+\min(\M2,\N2+\min(\N1,\M1))1({\M2\geq \N1})
\end{align}
where $1(\cdot)$ is indicator function.
\end{lemma}
\begin{IEEEproof}
If $\M2\geq \N1$ and assume transmitter 1 sends $d_1$ streams, transmitter 2
can send at most $\M2-\N1$ streams along the null space of $\bdH_{12}$ without
interfering receiver 1. Transmitter 2 can also send at most $\N1-d_1$ streams
along the row space of $\bdH_{12}$. Therefore user 2 can decode
$\min((\M2-\N1)+(\N1-d_1),\N2)$ streams without interfering receiver 1. If
$\N1 \geq \M2$ and assume transmitter 2 sends $d_2$ streams which interfere
receiver 1, transmitter 1 can send $\min(\N1-d_2,\M1)$ decodable streams to
receiver 1. Combining these two cases, we have the achievable DoF region shown
in this lemma.
\end{IEEEproof}

\begin{lemma}[Conversepart of \thmref{thm.zic.csit}] \label{lem.zic.outer}
The region given by \eqref{eq.zic.region1} and \eqref{eq.zic.region2} is a
valid outer bound for the two-user MIMO ZIC with CSIT.
\end{lemma}
\begin{IEEEproof} It is obvious that
adding antennas at the receiver will not shrink the DoF region. Hence, we can
add $\M1$ antennas to receiver 2 resulting an $(\M1,\N1,\M2,\M1+\N2)$ MIMO
FIC, and \eqref{eq.zic.region2} follows from Corollary 1 in \cite{jafa07}. The
outer bound of such a MIMO FIC is a valid outer bound of an
$(\M1,\N1,\M2,\N2)$ MIMO ZIC. Combining the trivial upper bound on
point-to-point system, we have this lemma.
\end{IEEEproof}

Based on \lemref{lem.zic.inner}, zero forcing at receiver is sufficient to
achieve the DoF region of ZIC when CSIT is available. The antenna mode
switching ability is not needed in this case. However, we shall see later that
such an ability is important for the case when CSIT is absent.

\section{Two-User MIMO ZIC and FIC without CSIT When Number of Modes $K \geq
N_1$ } \label{sec.kgeqn1} In this section, we describe the DoF region of
two-user ZIC and FIC without CSIT but with transmitter side reconfigurable
antennas. We deal with the case that $K$, the number of antenna modes is at
least equal to the $N_1$. The case $K<N_1$ will be dealt with in
\secref{sec.lessmodes}.

Based on the antenna number configuration, the achievability scheme of ZIC and
FIC without CSIT can be divided into two cases. In the first case, no
reconfigurable antenna is needed to achieve an DoF outer bound ---
reconfigurable antennas are not helpful (\secref{sec.noneed}). In the second
case, the outer bound can be achieved with enough transmit side antenna modes
(\secref{sec.needmode}): reconfigurable antennas enlarges the DoF region. Our
main results in this section are the following two theorems.

\begin{theorem}[ZIC with Enough Reconfigurable Antenna Modes]
\label{thm.zic.nocsit}
The DoF region of two-user MIMO Z interference channel without CSIT is
described by the following inequalities
\begin{align}
&d_i \leq \min(M_i,N_i), \quad i=1,2; \label{eq.zic.nocsit.0} \\
&d_1 \!+\!\frac{\min(\N1,\N2,\M2)}{\min(\N2,\M2)}d_2\!\leq\!
        \min(M_1\!+M_2,N_1).
        \label{eq.zic.nocsit.1}
\end{align}
if either one of the following is true:
\begin{enumerate}
\item[C1)] $M_1<N_1<\min(M_2, N_2)$ and transmitter one can switch among
        $N_1$ antenna modes, or
\item[C2)] $(M_1,N_1,M_2,N_2)$ do not satisfy the above condition. \hfill
\qed
\end{enumerate}
\end{theorem}

\begin{figure*}
\centering
\includegraphics[width=0.9\textwidth]{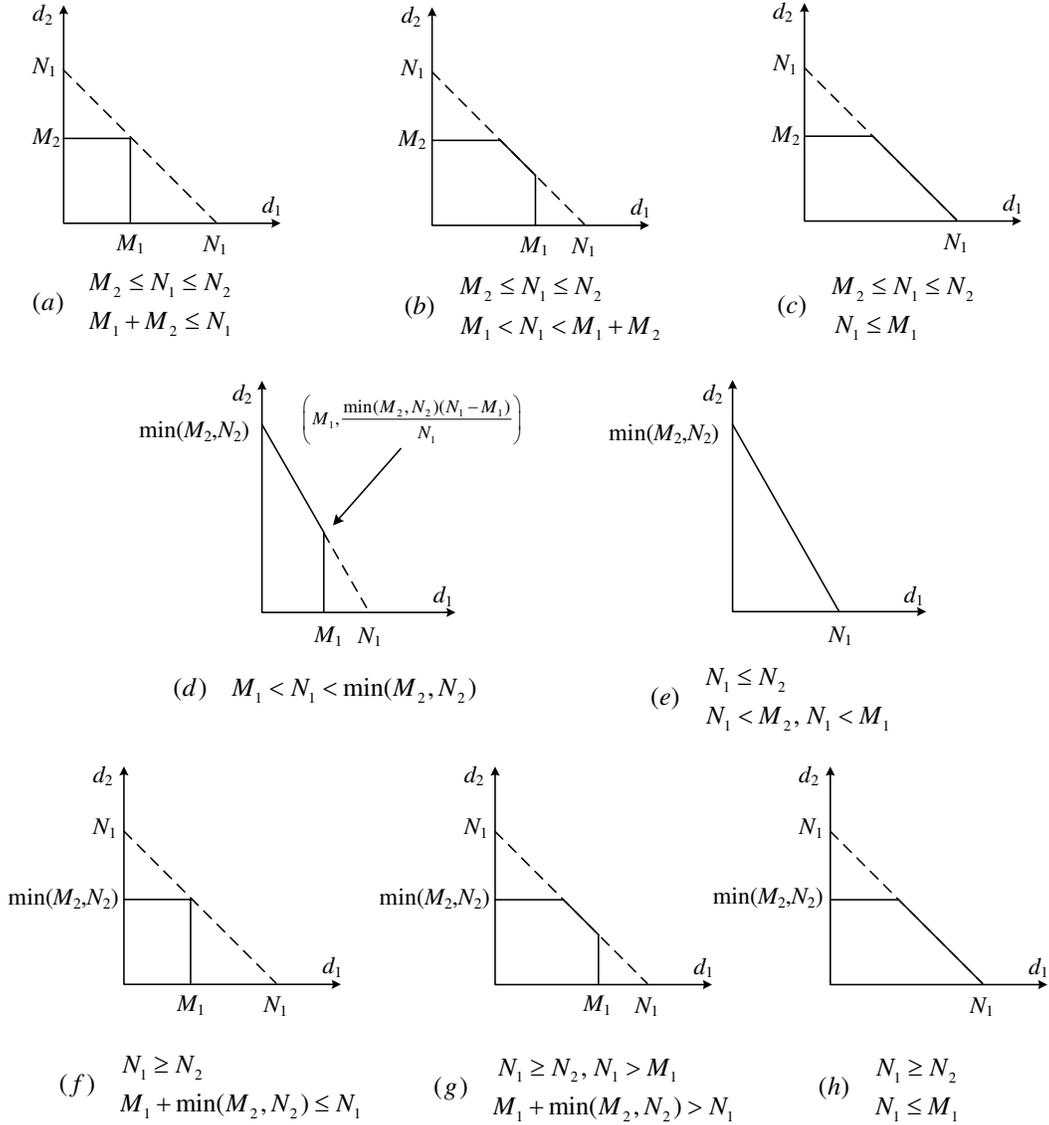}
\caption{DoF region of two-user MIMO ZIC without CSIT when number of antenna
modes $K\ge N_1$. Figures (a)--(e) are for the case $N_1\le N_2$; Figures
(f)--(h) are for the case $N_1\geq N_2$.}
\label{fig.dofzic}
\end{figure*}

The DoF region in \thmref{thm.zic.nocsit} is shown in \figref{fig.dofzic}.

\begin{theorem}[FIC with Enough Reconfigurable Antenna Modes]
\label{thm.fic.nocsit}
The DoF region of two-user MIMO full interference channel without CSIT is
described by the inequalities
\eqref{eq.nocsit.upper0}--\eqref{eq.nocsit.upper1} if any one of the following
is true:
\begin{enumerate}
\item[C1)] $M_1<N_1<\min(M_2, N_2)$ and transmitter one can switch among
        $N_1$ antenna modes, or
\item[C2)] $M_2<N_2<\min(M_1, N_1)$ and transmitter two can switch among
        $N_2$ antenna modes, or
\item[C3)] $(M_1,N_1,M_2,N_2)$ are not one of the two above cases. \hfill
\qed
\end{enumerate}
\end{theorem}

\subsection{Converse part} We first prove the converse part of the two
theorems.

\begin{lemma}[Converse part of \thmref{thm.fic.nocsit}]
\label{lem.th3converse}
The outer bound of DoF region of two-user MIMO full interference channel given
in \eqref{eq.nocsit.upper0}--\eqref{eq.nocsit.upper1} is still valid when
either or both transmitters are using antenna mode switching.
\end{lemma}
\begin{IEEEproof}
The outer bound \eqref{eq.nocsit.upper.m} has been derived based on the
assumption that the rows of $\bdH_{12}$ and those of $\bdH_{22}$ are
statistically equivalent \cite{hjsv09,zhgu09c}. Similarly, the outer bound
\eqref{eq.nocsit.upper1} has been derived based on the assumption that the
rows of $\bdH_{11}$ and those of $\bdH_{21}$ are statistically equivalent.
These assumptions are not affected by antenna mode switching at either or both
transmitters. Hence, the DoF outer bound is still valid.
\end{IEEEproof}

\begin{lemma}[Converse part of \thmref{thm.zic.nocsit}]
\label{thm.nocsit.zDoF} The outer bound of degrees of freedom region
of two-user MIMO Z interference channel without CSIT can be given as when
transmitter one has the antenna mode switching ability\begin{align} &d_i \leq
\min(M_i,N_i), \quad i=1,2; \label{eq.outer.z0} \\ &d_1
\!+\!\frac{\min(\N1,\N2,\M2)}{\min(\N2,\M2)}d_2\!\leq\!
        \min(M_1\!+M_2,N_1).
        \label{eq.outer.z}
\end{align}
\end{lemma}

\begin{IEEEproof}
This is the direct result of \cite[Theorem 1]{zhgu09c} as in
\eqref{eq.nocsit.upper0}--\eqref{eq.nocsit.upper1}, by noticing that there is
no interference from transmitter 1 to receiver 2 hence
\eqref{eq.nocsit.upper1} is not longer needed. The antenna switching at
transmitter one does not affect the upper bound, for the same reason stated in
\lemref{lem.th3converse}.
\end{IEEEproof}

\subsection{Achievability: when antenna mode switching is not needed}
\label{sec.noneed}
In this section, we prove the achievability part for Case C2) of
\thmref{thm.zic.nocsit} and Case C3) of \thmref{thm.fic.nocsit}. Achievability
for the remaining cases are left to \secref{sec.needmode}.

\begin{lemma} \label{lem.znocsit.n1ln2}
For the two-user MIMO Z interference channel without CSIT, when $N_1 \ge \N2$,
\eqref{eq.outer.z} is achievable by zero forcing.
\end{lemma}
\begin{IEEEproof} When $N_1 \ge \N2$, the corresponding outer regions are
shown in \figref{fig.dofzic} (f)--(h). Noticing that \eqref{eq.outer.z} is
reduced to \( d_1+d_2\leq\min(M_1+M_2,N_1) \), zero forcing is sufficient to
achieve the outer bound.
\end{IEEEproof}

\begin{lemma} \label{lem.nocsit.zic.min} When CSIT is
absent, the DoF outer region given by \lemref{thm.nocsit.zDoF} of a two-user
MIMO $(\M1,\N1,\M2,\N2)$ ZIC is the same as that of an
$(\M1,\N1,\min(\M2,\N2),\min(\M2,\N2))$ ZIC.
\end{lemma}
\begin{IEEEproof} We give the proof case by case. It is trivial that
when $\M2\leq\N2$ reducing the number of antennas at receiver 2 to $\M2$ will
not shrink the DoF region. When $\M2>\N2$, we can further consider two
sub-cases: $\N2 \geq \N1$ and $\N2 < \N1$.
\begin{enumerate}
\item When $\M2>\N2 \geq \N1$, corresponding to \figref{fig.dofzic} (d) and
(e), the DoF bound \eqref{eq.outer.z} becomes
$\frac{d_1}{\N1}+\frac{d_2}{\N2}\leq 1$. Hence the DoF outer region is the
same as an $(\M1,\N1,\N2,\N2)$ ZIC.
 \item When $\M2>\N2$ and $\N2 < \N1$, the DoF bound \eqref{eq.outer.z} becomes
$d_1+d_2\leq \min(\M1+\M2,\N1)$. Hence, if $\M1\geq\N1-\N2$, which implies
$\M1+\min(\M2,\N2)\geq\N1$, the DoF outer region is a pentagon or a tetragon;
see \figref{fig.dofzic} (g) and (h). Otherwise, it is a square, see
\figref{fig.dofzic} (f). One can show that the region is the same as that of
an $(\M1,\N1,\N2,\N2)$ ZIC.
\end{enumerate}
Hence, the lemma holds.
\end{IEEEproof}

We also have the following lemma regarding the relationship between DoF
regions of ZIC and FIC.

\begin{lemma}
\label{lem.ficziceq} When $\N1\le \N2$, the MIMO ZIC and FIC have
the same DoF regions. Any encoding scheme that is DoF optimal for one channel
is also DoF optimal for the other.
\end{lemma}
\begin{IEEEproof} Any point in the FIC is also trivially achievable in the ZIC
because user 2's channel is interference free. Conversely, any point
achievable in the ZIC region, is also achievable in FIC. This is based on the
fact that the channels are statistically equivalent at both receivers. If
receiver 1 can decode user 1's message, then receiver 2, having at least as
many antennas, must also be able to decode the same message. Receiver 2 can
then subtract the decoded message, which renders the resulting channel the
same as in the ZIC.
\end{IEEEproof}

Due to \lemref{lem.ficziceq}, we can translate all achievability schemes from
FIC to ZIC and vice versa when $\N1\leq \N2$. Therefore the achievability
schemes in \cite{hjsv09} for FIC when $N_1\le N_2$ and $M_1\ge N_1$ can be used
for ZIC. Therefore, the achievability part for Case C2) of
\thmref{thm.zic.nocsit} is complete.

For the FIC, the achievability for the case $N_1\le N_2$, except when
$M_1<N_1<\min(M_2,N_2)$, is shown in \cite{hjsv09}. When $N_1\ge N_2$, we can
swap the indices of the two users, so that except for the Cases C1) and C2)
the achievability scheme is known for FIC.

\subsection{Achievability: with antenna mode switching when $K\ge M_1N_1$}
\label{sec.needmode}

In this subsection, we prove a weaker version of the achievability for Case
C1) of \thmref{thm.zic.nocsit} and Cases C1) and C2) of
\thmref{thm.fic.nocsit}. Namely, we assume that the number of antenna modes
available is $K\ge M_1N_1$. The scheme is simpler in this case, and the
achievability scheme for the case $K=N_1$ will be built upon this case.

Based on \lemref{lem.nocsit.zic.min} and \lemref{lem.ficziceq}, we only
consider the two-user MIMO ZIC with $\M1 < \N1 < \M2=\N2$ to prove the Cases
C1) for both theorems. Case C2) of \thmref{thm.fic.nocsit} is the Case of C1)
with user indices swapped. Therefore, we want to show that the following DoF
pair is achievable for ZIC with $K_1=M_1N_1$ modes:
\begin{align}(d_1,
d_2)=\left(\M1,\frac{\M2(\N1-\M1)}{\N1}\right).
\end{align}

We first notice that this point cannot be achieved by zero forcing over one
time instant. This is because using zero forcing if transmitter 1 sends $\M1$
streams, transmitter 2 can only send $\N1-\M1$ streams without interfering
receiver 1. If transmitter 2 sends more streams, the desired signal and
interference are not separable at receiver 1 as transmitter 2 does not know
channel state information so it cannot send streams along the null space of
$\bdH_{12}$. A simple example is the $(1,2,3,3)$ case, where the outer bound
gives us $(d_1,d_2)=(1,1.5)$, which is not achievable via zero forcing over
one time slot. We make the assumption that the channel $\bdH_{12}$ stays the
same for at least $\N1$ time slots. It is sufficient to show that
$(\M1\N1,\M2(\N1-\M1)) $ streams can be achieved in $\N1$ time slots.

We first develop the beamforming and nulling design by assuming that there are
$N_1M_1$ antenna modes available at transmitter 1 such that it can use
different antenna modes in different slots to create channel variation. We
will further show that the resultant beamforming and nulling design still work
even if there are only $N_1$ modes available.

Here and after, we use tilde notation to indicate the time expansion signals,
where the number of slots of time expansion signals shall be clear within the
context. The time expansion channel between transmitter 1 and receiver 1 in
$\N1$ time slots is
\begin{align}
\tilde \bdH_{11}\!\!=\!\!\left[\! \begin{array}{ccccc}
\!\bdH_{11}(1) & \bdzeros & \bdzeros  & \bdzeros \\
\bdzeros & \!\bdH_{11}(2) & \bdzeros  & \bdzeros \\
\vdots & \vdots & \ddots & \vdots \\
\bdzeros & \bdzeros & \bdzeros  &\!\bdH_{11}(\N1)\! \\
\end{array} \right]_{\!\N1^2\! \times\! \N1\M1}\nonumber
\end{align}
and the channel between transmitter 2 and receiver 1 is
\begin{align}
\tilde \bdH_{12}=\bdI_{\N1}\otimes \bdH_{12}(1)
\end{align}
as transmitter 2 does not create channel variation. We will use precoding at
transmitter 2 only and nulling at receiver 1 only. Let $\tilde \bdP$ be the
transmit beamforming matrix at transmitter 2 and $\tilde \bdQ$ be the nulling
matrix at receiver 1. We propose to use the following structures for them
\begin{align}
\tilde \bdP_{\M2\N1 \times \M2(\N1-\M1)}&=\bdP_{\N1 \times
        (\N1-\M1)} \otimes \bdI_{\M2} \label{eq.Ptilde}\\
\tilde \bdQ_{\M1\N1 \times \N1^2}&=\bdQ_{\M1 \times \N1} \otimes
\bdI_{\N1}. \label{eq.Qtilde}
\end{align}

The received signal at receiver 1 can be written as
\begin{align}
\tilde \bdy_{1}=\tilde \bdH_{11} \tilde \bdx_1+\tilde \bdH_{12}
\tilde \bdP \tilde \bdx_2 +\tilde \bdz_1
\end{align}
where $ \tilde \bdx_1$ is a length $\M1\N1$ vector, and $ \tilde \bdx_2$ is a
length $\M2(\N1-\M1)$ vector. After applying nulling matrix $\tilde \bdQ$, we
have
\begin{align}
\tilde \bdQ\tilde \bdy_{1}=\underbrace{\tilde \bdQ\tilde
\bdH_{11}}_{\tilde\bdA} \tilde \bdx_1+\underbrace{\tilde \bdQ\tilde
\bdH_{12} \tilde \bdP}_{\tilde\bdB}  \tilde \bdx_2 +\tilde
\bdQ\tilde \bdz_1.  \label{eq.rx1null}
\end{align}
To achieve the degrees of freedom $(\M1\N1,\M2(\N1-\M1))$ for both users, it
is sufficient to design our $\tilde \bdP$ and $\tilde \bdQ$ to satisfy the
following conditions simultaneously
\begin{enumerate}
\item $\rank(\tilde\bdA)=\M1\N1$,
\item $\rank(\tilde \bdP)=\M2(\N1-\M1)$,
\item $\tilde \bdB=\bdzeros$.
\end{enumerate}
The second condition can be easily satisfied. Because $\rank(\tilde
\bdP)=\rank(\bdP) \rank(\bdI_{\M2})$, we only need to design $\bdP$ such that
$\rank(\bdP)=\N1-\M1$. As to the third condition, notice that
\begin{align*}
\tilde \bdB&=(\bdQ\otimes \bdI_{\N1})(\bdI_{\N1}\otimes \bdH_{12}(1))
	(\bdP \otimes \bdI_{\M2})\\
&=(\bdQ\bdI_{\N1}\bdP)\otimes (\bdI_{\N1}\bdH_{12}(1) \bdI_{\M2})\\
&=(\bdQ\bdP)\otimes \bdH_{12}(1).
\end{align*}
It is therefore sufficient (and also necessary) to have $\bdQ\bdP=\bdzeros$.
Then the key is to find a $\bdQ$ such that the equivalent channel of user 1
after nulling
\begin{align}\label{eq.Atilde}
\tilde\bdA=(\bdQ \otimes \bdI_{\N1} )\tilde\bdH_{11}
\end{align}
has full rank $\M1\N1$ with probability 1. The matrix $\tilde \bdA$ is of size
${\M1\N1 \times \M1\N1}$ and has the following structure
\begin{align}
\!\!\!\tilde \bdA \!\! =\!\!\left[ \begin{array}{ccccc}
\!\!\!q_{11}\bdH_{11}(1)\!\!\! & \!\!\!q_{12}  \bdH_{11}(2)\!\!\! &\!\!\! \cdots
& \!\!\!q_{1N_1}  \bdH_{11}(\N1)\!\!\!\\
\!\!\!q_{21}\bdH_{11}(1)\!\!\! & \!\!\!q_{12}  \bdH_{11}(2)\!\!\! &\!\!\! \cdots
& \!\!\!q_{2N_1}  \bdH_{11}(\N1)\!\!\!\\
\vdots & \vdots & \!\!\!\ddots & \vdots \\
\!\!\!q_{\M11}\bdH_{11}(1) & \!\!\!q_{\M12}  \bdH_{11}(2) &
\!\!\!\cdots  & \!\!\!q_{M_{1}N_1}  \bdH_{11}(\N1)\!\!\!
\end{array} \right]. \nonumber
\end{align}

To show that $\tilde \bdA$ has full rank, we need the following lemma, which
is known before, and a proof of it can be found in e.g., \cite{jist01}.

\begin{lemma}
\cite[Lemma 2]{jist01}\label{lem.fullrank} Consider an analytic function
$h(\bdx)$ of several variables $\bdx=[x_1,\dots,x_n]^T \in\mathbb{C}^n$. If
$h$ is nontrivial in the sense that there exists $\bdx_0\in\mathbb{C}^n$ such
that $h(\bdx_0)\neq 0$, then the zero set of $f(\bdx)$ \(
\mathrm{Z}:=\{\bdx\in \mathbb{C}^n | h(\bdx)=0 \} \) is of measure (Lebesgue
measure in $\mathbb{C}^n$) zero. \hfill \qed
\end{lemma}

Because the determinant of $\tilde \bdA$ is an analytic polynomial function of
elements of $\bdH_{11}(t), t=1,\dots,\N1$, we only need to find a specific
pair of $\bdQ$ and $\bdH_{11}(t), t=1,\dots,\N1$, such that $\tilde \bdA$ is
full rank. We propose the following:
\begin{align}\label{eq.Q}
 \bdQ=[\vand_{N_1}(1, \omega_{\N1},\dots,\omega^{M_{1}-1}_{\N1})]^T,
\end{align}
where $\omega_{N_1}\bydef \exp(-j{2\pi}/{N_1})$.

Let $\omega\bydef \exp(-j{2\pi}/{N_1^2})$. Take the realizations of
$\bdH_{11}(t)$, $t=1,\dots, N_1$, as
\begin{align}
 \bdH_{11}(t)=\vand_{N_1}(\omega^{t-1},\omega^{N_{1}+t-1},\dots,
 	\omega^{(M_{1}-1)N_{1}+t-1}).
\end{align}

It can be verified that for such choices of $\bdQ$ and $\bdH_{11}(t)$, $\tilde
\bdA$ is a Vandermonde matrix:
\begin{align*}
 \tilde
 \bdA=\vand_{M_1N_1}(1,\omega^{N_{1}},\dots,\omega^{(M_{1}-1)N_{1}},\omega^{1},
 	\omega^{N_{1}+1},\dots,&\omega^{(M_1-1)N_{1}+1},\\
 & \dots, \omega^{N_1-1},  \omega^{2N_1-1},\dots,\omega^{M_1N_{1}-1}
 ),
\end{align*}
hence of full rank. We also notice that $\tilde \bdA$ is a leading principal
minor of a permuted fast Fourier transform (FFT) matrix with size $\N1^2\times
\N1^2$. The permutation is as follows: Index the columns of an FFT matrix
$0,1,\dots, N_1^2-1$, and then permute them in an order shown below:
\begin{align*}
(0,N_1,2N_1,\dots,(M_1-1)N_1),\lb (1,N_1+1,2N_1+1, \dots,
(M_1-1)N_1+1),...
\end{align*}

Based on \lemref{lem.fullrank}, if we choose the nulling matrix using $\bdQ$
as specified in \eqref{eq.Q}, $\tilde\bdA$ has full rank almost surely. One
choice of the corresponding $\bdP$ matrix with respect to \eqref{eq.Q} is the
following
\begin{align} \label{eq.P}
 \bdP=\vand_{N_1}(\omega_{\N1}^{-M_1},\omega_{\N1}^{-(M_1+1)},\dots,
 	\omega^{-(N_1-1)}_{\N1}),
\end{align}
which is orthogonal to $\bdQ$. This completes the achievability part under
conditions in Case C1) of \thmref{thm.zic.nocsit} and Cases C1) and C2) of
\thmref{thm.fic.nocsit}, but with $K\ge M_1N_1$.

\subsection{Achievability: with antenna mode switching when $K=N_1$}
\label{sec.cyclic}

Assuming there are $N_1$ modes available at transmitter 1 and denote these
channel vectors between receive antennas of user 1 and the $i$th mode as
$\bdh_i, 1\leq i\leq \N1$ and let $\hat \bdH_{N_1\times
N_1}=[\bdh_1,\bdh_2,\dots,\bdh_{N_1}]$. We choose the antenna modes to be
switched cyclically:
\begin{align}
\bdH_{11}(1)&=[\bdh_1,\bdh_2,\dots,\bdh_{M_1}],\label{eq.modes.pt0} \\
\bdH_{11}(2)&=[\bdh_2,\bdh_3,\dots,\bdh_{M_1+1}],\\
& \vdots\nonumber\\
\bdH_{11}(N_1)&=[\bdh_{N_1},\bdh_1,\dots,\bdh_{M_1-1}]. \label{eq.modes.pt1}
\end{align}
We want to show that under this switching pattern, the equivalent channel
$\tilde \bdA$ in \eqref{eq.Atilde} between transmitter one and receiver one
after nulling, is still full rank. To show this, indexing the columns of
$\tilde\bdA$ in \eqref{eq.Atilde} as $0,1,\dots,\N1 \M1-1$, we then permute
and group the columns of $\tilde\bdA$ in the following way:
\begin{align*}
(0,M_1,2M_1,\dots,(M_1-1)N_1),\lb (1,M_1+1,2M_1+1, \dots,
(M_1-1)N_1+1),...
\end{align*}
Denote the permutation result as $\tilde\bdA '$ and it can be expressed as
\begin{equation}\label{eq.eqch.u1}
\tilde \bdA'  =\begin{bmatrix}\hat \bdH & \hat \bdH & \cdots & \hat \bdH\\
 \bdG \hat \bdH & \omega_{N_1}^{-1} \bdG \hat \bdH & \cdots &
 	\omega_{N_1}^{-M_1} \bdG \hat \bdH \\
\vdots & \vdots & \ddots & \vdots \\
\bdG^{M_1} \hat \bdH  & (\omega_{N_1}^{-1}\bdG)^{M_1} \hat \bdH & \cdots
	& (\omega_{N_1}^{-M_1}\bdG)^{M_1} \hat \bdH  \\
\end{bmatrix}, \nonumber
\end{equation}
where $\bdG$ is a size $\N1\times \N1$ diagonal matrix and can be expressed as
\begin{align}
\bdG=\diag(1,\omega_{N_1},\omega_{N_1}^2,\dots,\omega_{N_1}^{N_1-1} ).
\end{align}
Notice that $\tilde \bdA'= \bdR (\bdI_{M_1}\otimes \hat\bdH)$,
where
\begin{align}
\bdR  =\begin{bmatrix} \bdI_{N_1} & \bdI_{N_1} & \cdots & \bdI_{N_1}\\
 \bdG & \omega_{N_1}^{-1} \bdG & \cdots & \omega_{N_1}^{-M_1} \bdG  \\
\vdots & \vdots & \ddots & \vdots\\
\bdG^{M_1} & (\omega_{N_1}^{-1}\bdG)^{M_1} & \cdots &
	(\omega_{N_1}^{-M_1}\bdG)^{M_1}   \\
\end{bmatrix}. \nonumber
\end{align}
Recall $\omega_{N_1}= \exp(-j{2\pi}/{N_1})$. To show $\tilde\bdA$ is full
rank, it is necessary to show $\bdR$ is full rank as $\bdI_{M_1}\otimes
\hat\bdH$ is full rank with probability 1. It can be verified that via row and
column permutations $\bdR$ can be changed to a block diagonal matrix with the
$i$th block being
\begin{align}
\vand_{M_1}( \omega_{N_1}^i,\omega_{N_1}^{i-1},\cdots,\omega_{N_1}^{i-M_1+1}),
\end{align}
which is full rank due to Vandermonde structure. Hence $\bdR$ is full rank. It
follows that $\bdA$ is full rank with probability 1. This completes the
achievability part under conditions in Case C1) of \thmref{thm.zic.nocsit} and
Cases C1) and C2) of \thmref{thm.fic.nocsit} for $K=N_1$.

\subsection{Discussion} \label{sec.dis}

\subsubsection{Frequency domain interpretation} We note that the matrix
$[\bdQ^\dg,\; \bdP]$ is an inverse FFT (IFFT) matrix in our construction
\eqref{eq.Ptilde}, \eqref{eq.Qtilde}, \eqref{eq.Q} and \eqref{eq.P}. This
observation yields an interesting frequency domain interpretation of our
construction. The signal of user 2 is transmitted over frequencies
corresponding to the last $\N1-\M1$ columns of an IFFT matrix, whereas the
first user's signal is transmitted on all frequencies. Due to the antenna mode
switching at transmitter 1, the channel between transmitter 1 and receiver 1
is now time-varying and we manually introduce frequency spread. User 1's
signal is spread from one frequency bin to all the frequencies while user 2's
signal remains in the last $\N1-\M1$ frequency bins. Therefore the signal in
the first $\M1$ bins is interference free, which can be used to decode user
1's message. The nulling matrix applied at receiver 1 has a projection
explanation as well. Left multiplying the left and right hand sides of
\eqref{eq.rx1null} with $\tilde \bdQ^\dg$ yields
\begin{align*}
\tilde \bdQ^\dg\tilde \bdQ\tilde \bdy_{1}&= \tilde \bdQ^\dg\tilde \bdQ\tilde
	\bdH_{11} \tilde \bdx_1 +\tilde \bdQ^\dg\bdQ\tilde \bdz_1\\
&=((\bdQ^\dg\bdQ)\otimes\bdI_{N_1})(\tilde \bdH_{11}\tilde \bdx_1+\tilde\bdz_1),
\end{align*}
where $\bdQ^\dg\bdQ$ is the frequency domain projection matrix. We can see
that the signal of user 1 is projected from $\N1$ frequencies to the first
$\M1$ frequencies.

\subsubsection{The Loss of DoF due to lack of CSIT} In two-user MIMO Z
interference channel without CSIT, losing CSIT will not shrink degrees of
freedom region if $\M2\leq\N1$ or $\M2>\N1\geq\N2+M_1$. For all the other
cases, the degrees of freedom region is strictly smaller when comparing with
the CSIT case.

This observation can be verified case by case. Notice that it is already shown
in \cite[Theorem 2]{hjsv09} that when $\M2\leq\N1\leq\N2$ absence of CSIT does
not reduce DoF region in two-user MIMO FIC. Because MIMO FIC and ZIC has the
same DoF region when $\N1\leq\N2$. We only need to consider the sub cases when
$N_1>N_2$, corresponding to (f)--(h) in \figref{fig.dofzic}.
\begin{enumerate}
\item If $M_2<N_1$ and $N_1>N_2$, the total DoF of MIMO ZIC is upper bounded
by $N_1$ due to \eqref{eq.zic.region2}, so the DoF region remains the same if
CSIT is absent.
\item If $\M2>\N1>\N2$, the DoF region of MIMO ZIC without CSIT is a square
only when $M_1+N_2\leq N_1$, same as that of ZIC with CSIT. Otherwise, the
maximum total DoF of ZIC with CSIT is $\min(M_2,N_1+N_2,\min(M_1,N_1)+N_2)$,
strictly larger than $N_1$ which is the maximum total DoF when CSIT is absent,
hence loss of CSIT reduces the DoF region. \end{enumerate}

\subsubsection{Alternative construction when $N_1/M_1=\beta\in\mathbb{Z}$}

When $\N1/\M1=\beta\in\mathbb{Z}$, instead of using the $\bdQ$ given in
\eqref{eq.Q} we can use the following $\bdQ_{\M1\times\N1}=\bdI_{\M1} \otimes
\bdones_\beta^T.$ We need to show that this $\bdQ$ matrix will lead to a full
rank $\tilde \bdA$. This can be achieved by choosing $\tilde \bdH_{11}$ such
that it can be decomposed as $\tilde \bdH_{11}=\bdI_{M_1}\otimes\tilde
\bdH_{11}'$, where
\begin{align}
\tilde \bdH_{11}'\!\!=\!\!\left[\! \begin{array}{ccccc}
\!\bdH_{11}(1) & \bdzeros & \dots  & \bdzeros \\
\bdzeros & \!\bdH_{11}(2) & \dots & \bdzeros \\
\vdots & \vdots & \ddots & \vdots \\
\bdzeros & \bdzeros & \dots &\!\bdH_{11}(\beta)\! \\
\end{array} \right]_{\!\N1\beta\! \times\! \N1}.\nonumber
\end{align}
For this $\tilde \bdH_{11}$
\begin{align*}
\tilde\bdA&=(\bdI_{M_1}\otimes \bdones_\beta^T \otimes \bdI_{\N1} )
	(\bdI_{M_1}\otimes\tilde \bdH_{11}')\\
&=\bdI_{M_1}\otimes ((\bdones_\beta^T \otimes \bdI_{\N1})\tilde \bdH_{11}'),
\end{align*}
which has full rank. For this choice of $\bdQ$, we only use $\beta M_1=N_1$
antenna modes in $\N1$ time slots.

Therefore, for the two-user MIMO ZIC and FIC when $\M1<\N1<\min(\M2,\N2)$ and
$\N1/\M1=\beta\in\mathbb{Z}$, $\beta$ fold time expansion is enough to achieve
the DoF region. We remark that this can be viewed as the generalization of the
case we discussed in \secref{sec.needmode} for $\N1=\beta$ and $\M1=1$. In
fact $\bdones_\beta^T$ is the nulling matrix $\bdQ$ given in \eqref{eq.Q} when
$\N1=\beta,\M1=1$.

\subsubsection{Successive Decoding in ZIC} For the two-user MIMO FIC when
$\M1<\N1<\min(\M2,\N2)$ and CSIT is absent, we need block decoding at both
receivers in general, which introduces decoding delay. Successive interference
cancellation decoder can be used at receiver 2 to reduce decoding delay.
Taking the case $\N1/\M1=\beta\in\mathbb{Z}$ as an example, we can use $\beta$
fold time expansion and choose $\bdQ=\bdones_\beta^T$. The corresponding
$\bdP$ matrix is not necessary to be the last $\beta-1$ columns of an
$\beta\times \beta$ FFT matrix. The following $\bdP$ matrix still satisfies
the design constraint
\begin{align}
\bdP_{\beta \times (\beta-1)} =\begin{bmatrix}\bdI_{\beta-1} \\
\bdones_{\beta-1}^T
\end{bmatrix}.
\end{align}
Here, $\bdP$ has a nice structure. Every stream of user 2 can be decoded
immediately as they are interference free. For other cases where $\M1$ cannot
divide $\N1$, we can still find a $\bdQ$, $\bdP$ pair through numerical
simulation such that the upper diagonal parts of $\bdP$ are all zeros and
contain small number of nonzero entries. Such a beamforming matrix can
guarantee the immediate decoding of user 2's signal the interference only
comes from the streams already decoded .

\section{Two-User MIMO ZIC and FIC without CSIT When Number of Modes $K < N_1$
} \label{sec.lessmodes}

In this section, we will present our result for the $K<N_1$ case. The main
result of this section is the following theorem.

\begin{theorem}\label{thm.lessmodes}
When $\M1 < \N1 < \min(\M2,\N2)$ and the antennas of transmitter 1 can be
switched among $K$ antenna modes, where $K < N_1$, the DoF region of two-user
MIMO ZIC and FIC without CSIT is given by the following inequalities
\begin{align}
 d_i &\leq \min(M_i,N_i), \quad i=1,2; \label{eq.modes.outer.0} \\
 d_1+\frac{K}{\min(M_2,N_2)-(N_1-K)}d_2
        &\leq M_1+
            \frac{K(N_1-M_1)+(\min(M_2,N_2)-N_1)(K-M_1)  }
                {\min(M_2,N_2)-(N_1-K)}
 \label{eq.modes.outer.1}
\end{align}

The DoF region of FIC for $\M2 < \N2 < \min(\M1,\N1)$ can be obtained by
switching the two user indices.\hfill \qed
\end{theorem}

The method of proof is heavily based on that in \cite{yzdg10a}, to which the
reader is referred for several lemmas that will be used and their proofs. Some
notation that is used in this section are the following. We use tilde notation
to denote the time expanded signal \emph{over $L$ time slots} and $t\in [1,
L]$ is the index of the slot within one block. In general, by default, for a
vector $\bdx$, $\tilde \bdx=\vt (\bdx(1),\bdx(2),\cdots,\bdx(L))$ and for a
matrix $\bdV$, $\tilde \bdV=\diag(\bdV(1),\bdV(2),\cdots,\bdV(L))$. In
addition, for a time expanded vector $\tilde \bdx$, we use $\tilde \bdx^n$ or
$\{\tilde \bdx\}^n$ to denote a sequence of $n$ successive blocks of $\tilde
\bdx$: $\tilde \bdx^n=\vt (\bdx(1), \bdx(2), \ldots, \bdx(nL))$. Furthermore,
$\bdx(t)^n$ is the sequence of $\bdx(t)$ which contains all the vector $\bdx$
of the $t$th slot of all $n$ blocks:
\(
\bdx(t)^n=\vt(\bdx(t), \bdx(t+L), \ldots, \bdx(t+(n-1)L))
\).
Similar notation is defined for matrices as well. We use $\bdH$ denotes
$(\bdH_{11},\bdH_{12},\bdH_{21},\bdH_{22} )$, hence $\tilde \bdH^n$ denotes
all the channel matrices over $n$ blocks. In addition, for a random vector
$\bdx$, ${\bdx}_G$ is a corresponding CSCG vector that has the same covariance
matrix as $\bdx$.

\subsection{The Converse Part}

We prove the converse part of \thmref{thm.lessmodes} in the following. Recall
that for $\M1 < \N1 < \min(\M2,\N2)$, the proof is equivalent for both FIC and
ZIC. We will only show the proof for ZIC. To make the proof self-contained, we
will go through some similar steps as in \cite{yzdg10a}, but avoiding details.

The converse is developed based on blocking for every $L$ slots. In each
block, the channel $\bdH_{12},\bdH_{22}$ stay the same with the decomposition
$\bdH_{12}=\bdW_{12}\bdLambda_{12}\bdV_{12}^\dg$ and
$\bdH_{22}=\bdW_{22}\bdLambda_{22}\bdV_{22}^\dg$, whereas $\bdH_{11}$ is
time-varying among $L$ slots due to antenna mode switching at transmitter 1.
Transmitter 1 has $K$ modes with $K<N_1$ and it can adopt arbitrary switching
pattern. Let $\Hall$ be an $N_1 \times N_1$ full rank random matrix such that
$\Hall = [\bdh_1, \bdh_2,\cdots, \bdh_{N_1}]$ and $\bdh_i, 1\leq i\leq K$ is
the random vector channel between the $i$th antenna mode and receive antennas
of user 1. We introduce the fictitious vectors $\{\bdh_i, K+1\leq i\leq N_1\}$
to simplify the proof. We assume $\Hall$ is isotropic fading and \iid over
blocks of length $L$ each, where $L$ naturally satisfy $L\geq \lceil K/M_1
\rceil$. We denote the decomposition of $\Hall$ as
$\tdWall\tdLmdall\tdVall^\dg$.

Furthermore, let $\bdE(t)$ of size $N_1\times M_1$ denote the antenna mode
selection matrix for time $t$. Let $\bde_m, 1\leq m \leq N_1$ be the $m$th
column of $\bdI_{\N1}$. Let $i(t)$ denote the mode index selected by antenna
$i$ at time $t$. Then the $i$th column of $\bdE(t)$ is $\bde_{i(t)}$. We have
$\bdH_{11}(t)=\Hall\bdE(t)$.

At receiver 1, from Fano's inequality, we have
\begin{align}
nLR_1-\delta_{nL}\leq \mI (\tilde\bdy_1^n; \tilde\bdx_{1}^{n}|\tilde \bdH^n ).
\end{align}
where $\delta_{nL}\to 0$ as $n\to \infty$. Denote
\begin{align}
\tilde \bdr&= \tilde\bdH_{11} \tilde\bdx_{1G}
	+\tilde \bdH_{12}^\dg\tilde\bdx_{2}+\tilde \bdz_1\\
\tilde \bdr_1&=\tilde \bdW_{12}^\dg \tilde\bdH_{11} \tilde\bdx_{1G}+\tilde
	\bdV_{12}^\dg\tilde\bdx_{2}+\tilde \bdn_1
\end{align}
where $\tilde\bdn_1=\tilde\bdW_{12}^\dg\tilde \bdz_1$. Using \cite[Theorem
3]{yzdg10a}, which says that Gaussian input can reduce the mutual information
by at most an $\ologP$ quantity, and two uses of chain rule we have
\begin{align}
nLR_1&-n\,\ologP \nonumber \\
&\leq \mI (\tilde\bdr^n;\bdx_{1G}^{n}|\tilde \bdH^n )\\
&= \mI\left ( \tilde\bdr^n;\bdx_{1G}^{n}|\tilde\bdx_{2}^{n},\tilde \bdH^n\right)
	+\mI( \tilde\bdr^n;\tilde\bdx_{2}^{n}|\tilde \bdH^n)
	-\mI\left( \{\tilde \bdH_{12}^\dg\tilde\bdx_{2}+\tilde \bdz_1\}^{n};\tilde
\bdx^{n}_{2} | \tilde \bdH^n   \right).
\end{align}
Using \cite[Lemma 2]{yzdg10a}, we have
\begin{align}
\mI\left( \{\tilde \bdH_{12}^\dg\tilde\bdx_{2}+\tilde \bdz_1\}^{n};\tilde
\bdx^{n}_{2} | \tilde \bdH^n   \right)
&=\mI\left( \{\tilde \bdW_{12}\tilde \bdLambda_{12}\tilde
	\bdV_{12}^\dg\tilde\bdx_{2}+\tilde \bdz_1\}^{n};\tilde
\bdx^{n}_{2} | \tilde \bdH^n   \right) \\
&= \mI\left( \{\tilde \bdLambda_{12}\tilde \bdV_{12}^\dg\tilde\bdx_{2}
	+\tilde \bdn_1\}^{n};\tilde
\bdx^{n}_{2}  \tilde \bdH^n   \right) \\
&\geq \mI\left( \{\tilde \bdV_{12}^\dg\tilde\bdx_{2}+\tilde \bdn_1\}^{n};\tilde
	\bdx^{n}_{2} | \tilde \bdH^n   \right) -n\,\ologP,
\end{align}
and
\begin{align}
\mI( \tilde\bdr^n;\tilde\bdx_{2}^{n}|\tilde \bdH^n)&=\mI\left( \{\tilde
\bdW_{12}^\dg \tilde\bdH_{11} \tilde\bdx_{1G}+\tilde \bdLambda_{12}\tilde
\bdV_{12}^\dg\tilde\bdx_{2}+\tilde \bdn_1\}^{n};\tilde\bdx_{2}^{n} |\tilde
\bdH^n\right) \\ &\leq\mI\left( \tilde\bdr_1^n;\tilde\bdx_{2}^{n} | \tilde
\bdH^n \right) +n\,\ologP.
\end{align}
Hence $R_1$ can be further bounded as
\begin{align}
nLR_1&-n\,\ologP \nonumber \\
&\le \mI\left ( \tilde\bdr^n;\tilde\bdx_{1G}^{n}|\tilde\bdx_{2}^{n},
	\tilde \bdH^n\right )+\mI( \tilde\bdr_1^n;\tilde\bdx_{2}^{n}|\tilde \bdH^n)
	-\mI\left( \{\tilde \bdV_{12}^\dg\tilde\bdx_{2}+\tilde \bdn_1\}^{n};\tilde
\bdx^{n}_{2} | \tilde \bdH^n   \right). \label{eq.conv.r1}
\end{align}
As to receiver 2, using Fano's inequality and \cite[Lemma 2]{yzdg10a}, we have
\begin{align}
nLR_2-\delta_{nL}&\leq \mI (\tilde\bdy^n_2;\tilde\bdx_{2}^{n}|\tilde \bdH^n )\\
&= \mI\left( \{\tilde \bdW_{22}\tilde \bdLambda_{22}
	\tilde \bdV_{22}^\dg\tilde\bdx_{2}+\tilde \bdz_2\}^{n};\tilde\bdx^{n}_{2}
		| \tilde \bdH^n   \right) \\
&\leq  \mI\left( \{\tilde \bdV_{22}^\dg\tilde\bdx_{2}+\tilde \bdn_2\}^{n};\tilde
\bdx^{n}_{2} | \tilde \bdH^n   \right) +n\,\ologP,
\end{align}
where $\tilde\bdn_2=\tilde\bdW_{22}^\dg\tilde \bdz_2$. Hence
\begin{align}
nLR_2&-n\,\ologP\leq   \mI\left(\tilde \bdr_1^n;\tilde
\bdx^{n}_{2} | \tilde \bdH^n   \right)-\mI\left(\tilde \bdr_1^n;\tilde
\bdx^{n}_{2} | \tilde \bdH^n   \right)
	+\mI\left( \{\tilde \bdV_{22}^\dg\tilde\bdx_{2}
	+\tilde\bdn_2\}^{n};\tilde\bdx^{n}_{2}|\tilde\bdr_1^n, \tilde\bdH^n \right).
	\label{eq.conv.r2}
\end{align}
Notice that by using Gaussian input, the following inequalities hold
\begin{align}
\mI\left(\tilde\bdr_1;\tilde\bdx_{1G}^{n}|
	\tilde\bdx_{2}^{n},\tilde \bdH^n\right)
&\leq \E\log\left(\det(\bdI_{LN_1}
	+\frac{P}{M_1}\tilde \bdH_{11}\tilde \bdH_{11}^\dg ) \right)\\
&=nLM_1\logP+nL\ologP,  \label{eq.conv.pop1} \\
\mI\left(\tilde \bdr_1^n;\tilde
\bdx^{n}_{2} | \tilde \bdH^n   \right)&\leq n \E\log\left(\frac{\det(\bdI_{LN_1}
	+\frac{P}{M_2}\tilde \bdW_{12}\tilde \bdW_{12}^\dg
	+\frac{P}{M_1}\tilde \bdH_{11}\tilde \bdH_{11}^\dg )}{\det(\bdI_{LN_1}
	+\frac{P}{M_1}\tilde \bdH_{11}\tilde \bdH_{11}^\dg )}  \right)\\
&=nL(N_1-M_1)\logP+nL\ologP. \label{eq.conv.pop2}
\end{align}

Then let $n\rightarrow \infty$, multiply \eqref{eq.conv.r2} with some positive
scalar $\mu$, add it with \eqref{eq.conv.r1} and use \eqref{eq.conv.pop1},
\eqref{eq.conv.pop2}, we have the following inequality
\begin{align}
nL[R_1+uR_2-\ologP] \leq nL M_1\logP+\mu nL(\N1-\M1)\logP+\eta, \label{eq.r1ur2}
\end{align}
where $\mu$ is to be determined and
\begin{align}
\eta=&\mu\mI\left( \{\tilde \bdV_{22}^\dg\tilde\bdx_{2}+\tilde \bdn_2\}^{n};\tilde
\bdx^{n} _{2}| \tilde \bdH^n   \right)\!
	-\!\mI\left( \{\tilde \bdV_{12}^\dg\tilde\bdx_{2}
	+\tilde \bdn_1\}^{n};\tilde \bdx^{n} _{2}| \tilde \bdH^n   \right)\!
	+\!(1\!-\!\mu)\mI(\tilde \bdr_1^n;\tilde \bdx^{n}_{2}|\tilde\bdH^n).
	\label{eq.eta}
\end{align}
Divide \eqref{eq.r1ur2} by $nL\logP$ and let $P\rightarrow \infty$, we have the
following inequality on the DoF of two users
\begin{align}
d_1+\mu d_2\leq M_1+\mu(N_1-M_1) + \lambda,
\end{align}
where \[\lambda=\frac{1}{nL}\lim_{P\rightarrow \infty}\frac{\eta}{\logP}.\]

Recall that $\tilde \bdr_1=\tilde \bdW_{12}^\dg \tilde\bdH_{11}
\tilde\bdx_{1G}+\tilde \bdV_{12}^\dg\tilde\bdx_{2}+\tilde \bdn_1$ and
$\bdH_{11}(t)=\Hall\bdE(t)$. We define
\begin{align}
\tilde \bdr_2&=\tdLmdall^{-1}
\tdWall^\dg \tilde \bdW_{12}\tilde \bdV_{12}^\dg\tilde\bdx_{2}
	+ \tdVall^\dg\tilde \bdE   \tilde \bdx_{1G}+\tdLmdall^{-1}
	\tdWall^\dg \tilde \bdW_{12}\tilde \bdn_1\\
\tilde \bdr_3&=
\tdWall^\dg \tilde \bdW_{12}\tilde \bdV_{12}^\dg\tilde\bdx_{2}
	+ \tdVall^\dg\tilde \bdE   \tilde \bdx_{1G}+\tdLmdall^{-1}
	\tdWall^\dg \tilde \bdW_{12}\tilde \bdn_1\\
\tilde \bdr_4&=
	\tdVall\tdWall^\dg \tilde \bdW_{12}\tilde \bdV_{12}^\dg\tilde\bdx_{2}
	+ \tilde \bdE \tilde  \bdx_{1G}+\tdVall\tdLmdall^{-1}
	\tdWall^\dg \tilde \bdW_{12}\tilde\bdn_1\\
\tilde \bdr_5&=
	\tdVall\tdWall^\dg \tilde \bdW_{12}\tilde \bdV_{12}^\dg\tilde\bdx_{2}
	+ \tilde \bdE \tilde  \bdx_{1G}+\tilde\bdn_1\\
\tilde \bdr_6&=\tilde \bdV_{12}^\dg\tilde\bdx_{2}
	+ \tilde \bdE \tilde  \bdx_{1G}+\tilde\bdn_1 \label{eq.r6}
\end{align}
We have
\begin{align}
 \mI(\tilde \bdr_1^{n};\tilde\bdx^{n}_2| \tilde \bdH^n)
&=\mI(\tilde \bdr_2^{n};\tilde\bdx^{n}_2|\tilde \bdH^n) \label{eq.ap0}\\
&=\mI(\tilde \bdr_3^{n};\tilde\bdx^{n}_2|\tilde \bdH^n)+o(\logP)\label{eq.ap1}\\
&=\mI(\tilde \bdr_4^{n};\tilde\bdx^{n}_2|\tilde \bdH^n)+o(\logP)\label{eq.ap2}\\
&=\mI(\tilde \bdr_5^{n};\tilde\bdx^{n}_2|\tilde \bdH^n)+o(\logP)\label{eq.ap3}\\
&=\mI(\tilde \bdr_6^{n};\tilde\bdx^{n}_2|\tilde \bdH^n)+o(\logP), \label{eq.ap4}
\end{align}
where \eqref{eq.ap1} due to \cite[Lemma 2]{yzdg10a}; \eqref{eq.ap0} and
\eqref{eq.ap2} hold as $\tdWall\tdLmdall$ and $\tdVall$ are full rank square
matrices. \eqref{eq.ap3} holds as changing noise variance will not change the
DoF. \eqref{eq.ap4} is true because $\tdVall\tdWall^\dg \tilde \bdW_{12}\tilde
\bdV_{12}^\dg$ has the same distribution as $\tilde \bdV_{12}^\dg$ and
$\tdVall\tdWall^\dg \tilde \bdW_{12}$ is independent of $\tilde
\bdV_{12}^\dg$. To find the DoF order of $\mI(\tilde
\bdr_6^{n};\tilde\bdx^{n}_{2}| \tilde \bdH^n)$, we first notice that for each
slot $t$ in one block, $ \bdV_{12}^\dg$ can be divided into three parts:
$\bdV_{12,a}^\dg(t)$, $\bdV_{12,b}^\dg(t)$ and $\bdV_{12,c}^\dg$.

\begin{enumerate}
\item $\bdV_{12,a}^\dg(t)$ is of size $\M1\times \M2$ and consists of $M_1$
non-zero rows of $\bdE(t)\bdV_{12}^{\dg}$.
\item $\bdV_{12,c}^\dg$ is of size $(N_1-K)\times \M2$ and is the same for all
$1\leq t \leq L$. It consists of $N_1-K$ rows of $ \bdV_{12}^\dg$ that do not
appear in any $\bdV_{12,a}(t)^{\dg}, 1\leq t \leq L$.
\item $\bdV_{12,b}^\dg(t)$ is of size $(K-\M1)\times \M2$ and consists of
$K-\M1$ rows of $ \bdV_{12}^\dg$ that neither in $\bdE(t)\bdV_{12,a}(t)^{\dg}$
nor in $\bdV_{12,c}^\dg$.
\end{enumerate}

\noindent \textbf{Example:} Assume $\N1=5$, $\M1=2$, $L=6$, $K=4$ and
$\bdV_{12}=[\bdv_1,\bdv_2,\dots,\bdv_{N_1}]$ where $\bdv_i$'s are $\M2\times
1$ vectors. Assume $\bdE(t)$ is the following
\begin{align*}
\bdE(1)&=[\bde_1,\bde_2],\quad\bdE(2)=[\bde_1,\bde_3],\quad\bdE(3)
=[\bde_1,\bde_4],\\ \quad \bdE(4)&=[\bde_1,\bde_2],\quad
\bdE(5)=[\bde_2,\bde_4], \quad \bdE(6)=[\bde_2,\bde_3].
\end{align*}
We have
\begin{align*}
&\bdV_{12,a}^\dg(1)=[\bdv_1,\bdv_2]^\dg, \quad
 \bdV_{12,a}^\dg(2) =[\bdv_1,\bdv_3]^\dg, \quad
 \bdV_{12,a}^\dg(3)=[\bdv_1,\bdv_4]^\dg,\\
&\bdV_{12,a}^\dg(4)=[\bdv_1,\bdv_2]^\dg, \quad
 \bdV_{12,a}^\dg(5) =[\bdv_2,\bdv_4]^\dg, \quad
 \bdV_{12,a}^\dg(6)=[\bdv_2,\bdv_3]^\dg,\\
&\bdV_{12,b}^\dg(1)=[\bdv_3,\bdv_4]^\dg, \quad
 \bdV_{12,b}^\dg(2) =[\bdv_2,\bdv_4]^\dg, \quad
 \bdV_{12,b}^\dg(3)=[\bdv_2,\bdv_3]^\dg,\\
&\bdV_{12,b}^\dg(4)=[\bdv_3,\bdv_4]^\dg, \quad
 \bdV_{12,b}^\dg(5) =[\bdv_1,\bdv_3]^\dg, \quad
 \bdV_{12,b}^\dg(6)=[\bdv_1,\bdv_4]^\dg,\label{eq.i6}
\end{align*}
and $\bdV_{12,c}^\dg=\bdv_5^\dg$. Note that $\bdV_{12,c}^\dg$ remains the same
in one block of $L$ slots. \hfill\qed

Suppose receiver 1 receives $\bdr_6$ as in \eqref{eq.r6} and wants to decode
the message of $\bdx_2$ that goes through an equivalent channel
$\bdV_{12}^\dg$. Then $\bdV_{12,a}^\dg(t)$ are the directions of interference
from transmitter at time $t$, $\bdV_{12,b}^\dg(t)$ are those directions that
are temporarily interference-free at time $t$, and $\bdV_{12,c}^\dg$ are the
directions which are interference free for a whole block. The associated
noises of the those directions are similarly defined as
$\bdn_{2,a}(t),\bdn_{2,b}(t)$ and $\bdn_{2,c}(t)$.

To bound the DoF of $ \mI(\tilde \bdr_6^{n};\tilde\bdx^{n}_2| \tilde \bdH^n)$
of \eqref{eq.i6}, we define
\begin{align}
\bdV_{12,ab}^\dg(t)=\begin{bmatrix}\bdV_{12,a}^\dg(t) \\
\bdV_{12,b}^\dg(t) \\
\end{bmatrix}, \quad \bdn_{1,ab}(t)=\begin{bmatrix}\bdn_{1,a}(t) \\
\bdn_{1,b}(t) \\
\end{bmatrix},\\
\bdV_{12,bc}^\dg(t)=\begin{bmatrix}\bdV_{12,b}^\dg(t) \\
\bdV_{12,c}^\dg \\
\end{bmatrix}, \quad \bdn_{1,bc}(t)=\begin{bmatrix}\bdn_{1,b}(t) \\
\bdn_{1,c}(t) \\
\end{bmatrix},
\end{align}
and adopt the following notation for simplicity
\begin{align}
\va(t)&=\bdV_{12,a}^\dg(t)\bdx_2(t)+\bdx_{1G}(t)+\bdn_{1,a}(t)\\
\vb(t)&=\bdV_{12,b}^\dg(t)\bdx_2(t)+\bdn_{1,b}(t)\\
\vc(t)&=\bdV_{12,c}^\dg\bdx_2(t)+\bdn_{1,c}(t)\\
\vbc(t)&=\bdV_{12,bc}^\dg(t)\bdx_2(t)+\bdn_{1,bc}(t)
\end{align}
In addition, $\va(t)^n,\vb(t)^n,\vc(t)^n,\vbc(t)^n$ are sequences of
corresponding vectors of the $t$th slot over $n$ blocks. The collection of
$\va(1)^n, \va(2)^n,\dots \va( t)^n$ is denoted as $\{\va^{(1:t)}\}^n$. We
also define $\{\vb^{(1:t)}\}^n$ , $\{\vc^{(1:t)}\}^n$ and $\{\vbc^{(1:t)}\}^n$
similarly. Using the chain rule, we have
\begin{align}
\mI(\tilde \bdr_6^{n};\tilde\bdx_{2}^{n}| \tilde \bdH^n)
&=\mI\left(  \{\tilde \bdV_{12,c}^\dg\tilde\bdx_{2}+\tilde\bdn_{1,c}\}^{n};
	\tilde \bdx_{2}^{n} | \tilde \bdH^n \right)\notag \\
&\quad +\mI\left(\{\tilde \bdV_{12,b}^\dg\tilde\bdx_{2}+\tilde \bdn_{1,b}\}^{n};
	\tilde \bdx_{2}^{n} | \{\tilde \bdV_{12,c}^\dg\tilde\bdx_{2}
	+\tilde \bdn_{1,c}\}^{n}, \tilde \bdH^n  \right)\notag\\
&\quad +\mI\left(\{\tilde\bdV_{12,a}^\dg\tilde\bdx_{2}+\tilde\bdx_{1G}
	+\tilde\bdn_{1,a}\}^{n};\tilde \bdx_{2}^{n}
		|\{\tilde \bdV_{12,bc}^\dg\tilde\bdx_{2}
	+\tilde \bdn_{1,bc}\}^n, \tilde \bdH^n  \right) \label{eq.va,bc}
\end{align}

Now checking the second term in \eqref{eq.va,bc}, we notice that
\begin{align}
&\mI\left(  \{\tilde \bdV_{12,b}^\dg\tilde\bdx_{2}
	+\tilde \bdn_{1,b}\}^{n};\tilde \bdx_{2}^{n}
		|\{\tilde \bdV_{12,c}^\dg\tilde\bdx_{2}
+\tilde \bdn_{1,c}\}^{n}, \tilde \bdH^n  \right)\nonumber \\
&=\sum_{t=1}^L\mI\left(\vb(t)^n;\tilde
\bdx_{2}^{n} |\{\vb^{(1:t-1)}\}^n,  \tilde \bdy_c^n, \tilde \bdH^n   \right)
	\label{eq.mtd2.a}\\
&=\sum_{t=1}^L\mH\left(\vb(t)^n |\{\vb^{(1:t-1)}\}^n,  \tilde \bdy_c^n,
	\tilde \bdH^n   \right)-\mH\left(\vb(t)^n |\tilde \bdx_{2}^{n},
	\{\vb^{(1:t-1)}\}^n,  \tilde \bdy_c^n, \tilde \bdH^n   \right)
	\label{eq.mtd2.b}\\
&=\sum_{t=1}^L\mH\left(\vb(t)^n |\{\vb^{(1:t-1)}\}^n,  \tilde \bdy_c^n,
	\tilde \bdH^n   \right)-\mH\left(\vb(t)^n
	| \bdx_{2}(t)^{n},  \vb(t)^n, \tilde \bdH(t)^n   \right) \label{eq.mtd2.c}\\
&\leq\sum_{t=1}^L\mH\left(\vb(t)^n |  \vc(t)^n, \bdH(t)^n   \right)
	-\mH\left(\vb(t)^n | \bdx_{2}(t)^{n},  \vc(t)^n, \bdH(t)^n   \right)
	\label{eq.mtd2.d}\\
& = \sum_{t=1}^L\mI\left(\vb(t)^n ;
\bdx_{2}(t)^{n}|  \vc(t)^n, \bdH(t)^n   \right)\label{eq.mtd2.e}\\
&= \sum_{t=1}^L \left[\mI\left(\vb(t)^n,\vc(t)^n;\bdx_{2} (t)^{n}
	| \bdH(t)^n\right)-\mI\left(\vc(t)^n;
\bdx_{2} (t)^{n}|  \bdH(t)^n\right) \right]\label{eq.mtd2.f}\\
&\leq \sum_{t=1}^L  \left(\frac{N_1-M_1}{N_1-K}-1\right)\mI\left(\vc(t)^n;
\bdx_{2} (t)^{n}| \bdH(t)^n\right)  \label{eq.mtd2.g}\\
&\le nL(K-M_1)\logP + o(\logP) \label{eq.mtd2.h}
\end{align}
where:
\begin{itemize}
\item \eqref{eq.mtd2.a} and \eqref{eq.mtd2.f} follow by chain rule.
\item \eqref{eq.mtd2.b} and \eqref{eq.mtd2.e} are expressing mutual
information via entropy.
\item \eqref{eq.mtd2.c} holds as the second term is the entropy of noise when
conditioning on $\bdx_{2}(t)^{n}$.
\item \eqref{eq.mtd2.d} is based on the fact that conditioning reduces
entropy.

\item \eqref{eq.mtd2.g} follows by \cite[Lemma 3]{yzdg10a}.
\item \eqref{eq.mtd2.h} holds due to the fact that the DoF of an
$(N_1-K)\times M_2$ point-to-point MIMO channel is at most
$\min(N_1-K,M_2)=N_1-K$.
\end{itemize}

The third term in \eqref{eq.va,bc} can be bounded in a similar fashion. We
have
\begin{align}
&\mI\left(  \{\tilde \bdV_{12,a}^\dg\tilde\bdx_{2}
	+\tilde\bdx_{1G}+\tilde \bdn_{1,a}\}^{n};\tilde
\bdx_{2}^{n} | \{\tilde \bdV_{12,bc}^\dg\tilde\bdx_{2}
	+\tilde \bdn_{1,bc}\}^n, \tilde \bdH^n  \right) \nonumber\\
&=\sum_{t=1}^L\mI\left(\va(t)^n;\tilde \bdx_{2}^{n}
	|\{\va^{(1:t-1)}\}^n,  {\tilde \bdy}_{bc}^n, \tilde \bdH^n   \right)
	\label{eq.mtd3.a}\\
&=\sum_{t=1}^L\mH\left(\va(t)^n |\{\va^{(1:t-1)}\}^n,  {\tilde \bdy}_{bc}^n,
	\tilde \bdH^n   \right)
	-\mH\left(\va(t)^n |\tilde \bdx_{2}^{n}, \{\va^{(1:t-1)}\}^n,
	{\tilde \bdy}_{bc}^n, \tilde \bdH^n   \right) \label{eq.mtd3.b}\\
&=\sum_{t=1}^L\mH\left(\va(t)^n |\{\va^{(1:t-1)}\}^n,  {\tilde \bdy}_{bc}^n,
	\tilde \bdH^n   \right)
	-\mH\left(\va(t)^n | \bdx_{2}(t)^{n},  \vbc(t)^n, \tilde \bdH^n   \right)
	\label{eq.mtd3.c}\\
&\leq\sum_{t=1}^L\mH\left(\va(t)^n |  \vbc(t)^n, \bdH(t)^n   \right)
	-\mH\left(\va(t)^n | \bdx_{2}(t)^{n},  \vbc(t)^n, \bdH(t)^n   \right)
	\label{eq.mtd3.d}\\
& = \sum_{t=1}^L\mI\left(\va(t)^n ;
\bdx_{2}(t)^{n}|  \vbc(t)^n, \bdH(t)^n   \right)\label{eq.mtd3.e}\\
&\leq n \sum_{t=1}^L\mI\left(\bdy_{aG}(t);
\bdx_{2G}(t)|  \bdy_{bcG}(t), \bdH(t)   \right)\label{eq.mtd3.f}\\
&\leq nL\,\log \left( \det\left( \frac{P}{M_1}\bdI_{M_1}+\bdI_{M_1}
	+\frac{P}{M_2}\bdI_{M_1} \right) \det\left( \frac{P}{M_2}\bdI_{(N_1-M_1)}
	+\bdI_{(N_1-M_1)} \right) \right) \nonumber\\
& \qquad \qquad - nL\,\log \left( \det\left( \frac{P}{M_2}\bdI_{(N_1-M_1)}
	+\bdI_{(N_1-M_1)} \right) \det\left( \frac{P}{M_1}\bdI_{M_1}
	+\bdI_{M_1} \right) \right)  \label{eq.mtd3.g}\\
&= o(\logP) \label{eq.mtd3.h}
\end{align}
where:
\begin{itemize}
\item \eqref{eq.mtd3.a} follows by chain rule.
\item \eqref{eq.mtd3.b} and \eqref{eq.mtd3.e} are expressing mutual
information via entropy.
\item \eqref{eq.mtd3.c} holds as the second term is the entropy of noise when
conditioning on $\bdx_{2}(t)^{n}$.
\item \eqref{eq.mtd3.d} is based on the fact that conditioning reduces
entropy.

\item \eqref{eq.mtd3.f} and \eqref{eq.mtd3.g} follows by \cite[Lemma
3]{yzdg10a}, where the covariance matrix of $\bdx_{1G}(t)+\bdn_{1,a}(t)$ and
$\bdn_{1,bc}(t)$ are $\frac{P}{M_1}\bdI_{M_1}+\bdI_{M_1}$ and
$\bdI_{(N_1-M_1)}$, respectively. In addition, the optimal input of
$\bdx_{2}(t)$ is CSCG with covariance matrix $\frac{P}{M_2}\bdI_{(N_1-M_1)}$.

\end{itemize}
Substitute \eqref{eq.mtd2.h} and \eqref{eq.mtd3.h} in to \eqref{eq.va,bc}, we
have
\begin{align}
\mI(\tilde \bdr_1^{n};\tilde\bdx^{n}_2| \tilde \bdH^n)
 \leq\mI\left(  \{\tilde \bdV_{12,c}^\dg\tilde\bdx_{2}
 +\tilde \bdn_{1,c}\}^{n};\tilde \bdx_{2}^{n} | \tilde \bdH^n  \right)
 + nL(K-M_1)\logP + o(\logP)
\end{align}

Now we go back to \eqref{eq.eta}. Notice that if we choose
$\bdD=[\bdzeros_{N_1\times(\min(M_2,N_2)-N_1)},\bdI_{N_1}]$,
$(\bdD\bdV_{22}^\dag,\bdD\bdn_2)$ has the same distribution as
$(\bdV_{12}^\dag,\bdn_1)$ as both $\bdV_{22}$ and $\bdV_{12}$ are uniformly
distributed and $\bdV_{22}$ has no fewer columns than $\bdV_{12}$. (Please
refer to \cite[Sec.~IV-C2]{yzdg10a} for more details). We have the following
Markov chain:
\begin{align}\tilde \bdx_{2} \text{ --- }
 \tilde \bdV_{22}^\dg\tilde\bdx_{2}+\tilde \bdn_2 \text{ --- }
 \tilde \bdV_{12}^\dg\tilde\bdx_{2}+\tilde \bdn_1  \text{ --- }
 \tilde \bdV_{12,c}^\dg\tilde\bdx_{2}+\tilde \bdn_{1,c}.
\end{align}
Denote $J=\min(M_2,N_2)-(\N1-K)$. Let $\bdV_{22,a}$ contain the first $J$ rows
of $\bdV_{22}$, and $\bdn_{2,a}$ contain the first $J$ elements of $\bdn_2$.
We can bound $\eta$ as
\begin{align}
\eta&\leq \mu\mI\left( \{\tilde \bdV_{22}^\dg\tilde\bdx_{2}
	+\tilde \bdn_2\}^{n};\tilde \bdx_{2}^{n}
	|\{\tilde \bdV_{12,c}^\dg\tilde\bdx_{2}
	+\tilde \bdn_{1,c}\}^{n}, \tilde \bdH^n   \right)\nonumber \\
&\qquad -\mI\left( \{\tilde \bdV_{12}^\dg\tilde\bdx_{2}
	+\tilde \bdn_1\}^{n};\tilde \bdx^{n} _{2}
		|\{\tilde \bdV_{12,c}^\dg\tilde\bdx_{2}
	+\tilde \bdn_{1,c}\}^{n} , \tilde \bdH^n   \right)\nonumber\\
&\qquad+(1-\mu)nL(K-M_1)\logP + o(\logP)\\
&=\mu \mI\left( \{\tilde \bdV_{22,a}^\dg\tilde\bdx_{2}
	+\tilde \bdn_2\}^{n};\tilde \bdx_{2}^{n}
		|\{\tilde \bdV_{12,c}^\dg\tilde\bdx_{2}
	+\tilde \bdn_{1,c}\}^{n}, \tilde \bdH^n   \right)\nonumber \\
&\qquad-\mI\left( \{\tilde \bdV_{12,ab}^\dg\tilde\bdx_{2}
	+\tilde \bdn_1\}^{n};\tilde \bdx^{n} _{2}
		|\{\tilde \bdV_{12,c}^\dg\tilde\bdx_{2}
	+\tilde \bdn_{1,c}\}^{n} , \tilde \bdH^n   \right)\nonumber\\
&\qquad+(1-\mu)nL(K-M_1)\logP + o(\logP) \label{eq.chooseu}
\end{align}
Notice that the size of $\bdV_{12,ab}^\dg$ is $K\times\M2$. Based on
\cite[Lemma 3]{yzdg10a}, if we choose
\begin{align}
\mu=\frac{K}{J}
\end{align}
the difference of the first two mutual information terms of \eqref{eq.chooseu}
is at most in the order of $o(\logP)$ and we have
\begin{align}
\lambda\le\left(1-\frac KJ \right)(K-M_1)=\frac{(\min(M_2,N_2)-N_1)(K-M_1)}
	{\min(M_2,N_2)-(N_1-K)}
\end{align}
Recall that $d_1+\mu d_2\leq M_1+\mu(N_1-M_1) + \lambda$. We thus have the
outer bound on the sum DoF as shown in \eqref{eq.modes.outer.1} and the
proof of the converse part of \thmref{thm.lessmodes} is complete. \hfill \qed

\subsection{Achievability}

In order to show the achievability part of \thmref{thm.lessmodes}, we only
need to construct an achievable scheme for the corner point of the DoF region.
Without loss of generality, we assume that $M_2=\min(M_2,N_2)$; otherwise,
transmitter 2 can simply use $N_2$ transmit antennas. Since
$K(N_1-M_1)+(M_2-N_1)(K-M_1)=M_2(K-M_1)+M_1(N_1-K)$, it is sufficient to show
that the following DoF pair
\begin{align}
 (d_1,d_2)=(KM_1,M_2(K-M_1)+M_1(N_1-K)) \label{eq.dofpair.kleqn1}
\end{align}
can be achieved over $K$ slots with antenna mode switching at transmitter one
among $K$ modes. Similar to \secref{sec.cyclic}, we choose the mode switching
pattern as follows:
\begin{align*}
\bdE(1)&=[\bde_1,\bde_2,\dots,\bde_{M_1}],\\
\bdE(2)&=[\bde_2,\bde_3,\dots,\bde_{M_1+1}],\\
& \vdots\\
\bdE(K)&=[\bde_{K},\bde_1,\dots,\bde_{M_1-1}].
\end{align*}

We propose to use a generalization of the joint nulling and beamforming design
that is investigated in \secref{sec.needmode}. Unlike the frequency nulling
that has been used for $K=N_1$, this scheme requires that receiver 1 performs
nulling in both frequency and spatial domains. We hereby use two superscripts
$\spF$ and $\spS$ to indicate the matrices that associated with frequency
processing and spatial processing.

The generalized joint nulling and beamforming has the following structure:
\begin{align}
\tilde \bdQ&=\bdQ^\spF_{\M1 \times K} \otimes
\bdQ^\spS_{K \times \N1},\\
\tilde \bdP &= [\tilde \bdP_a, \tilde \bdP_b], \text{ where} \\
\tilde\bdP_a&=[\bdP_a^\spF]_{K \times
        (K-\M1)} \otimes [\bdP^\spS_a]_{\M2 \times \M2},\\
\tilde\bdP_b&=[\bdP_b^\spF]_{K \times M_{1}}
	\otimes [\bdP^\spS_b]_{\M2 \times(N_1-K)}.
\end{align}
 The received signal at receiver 1 can be written as
\begin{align}
\tilde \bdy_{1}=\tilde \bdH_{11} \tilde \bdx_1+\tilde \bdH_{12} [
\tilde  \bdP_a,\tilde \bdP_b ]\tilde \bdx_{2} +\tilde \bdz_1
\end{align}
where $ \tilde \bdx_1$ is a length $\M1K$ vector, and $ \tilde \bdx_{2}$ is a
length $M_2(K-M_1)+M_1(N_1-K)$ vector.

\begin{figure*}
\centering
\includegraphics[width=0.8\textwidth]{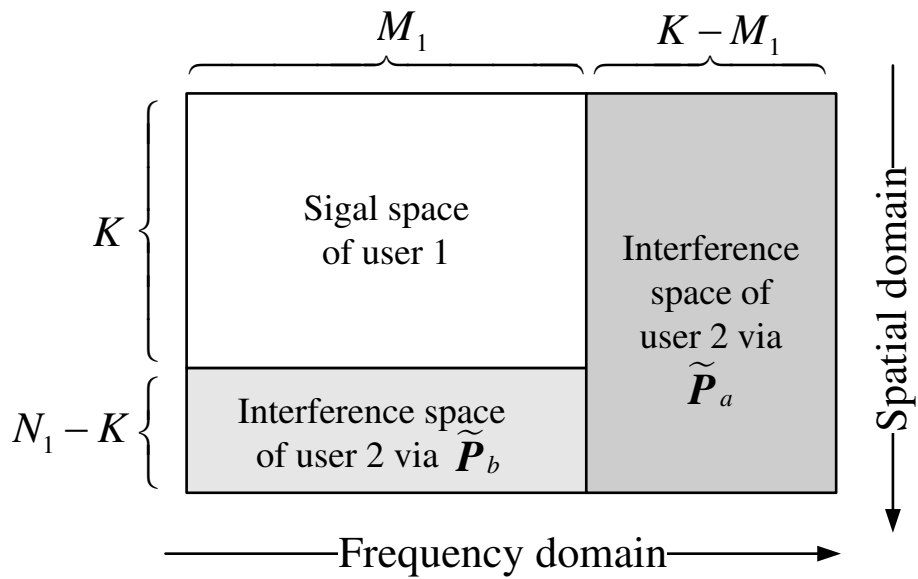}
\caption{Space-Frequency dimension allocation for the two users when $K<N_1$.}
\label{fig.spacefreq}
\end{figure*}

After applying nulling matrix $\tilde \bdQ$, we have
\begin{align}
\tilde \bdQ\tilde \bdy_{1}=\underbrace{\tilde \bdQ\tilde
\bdH_{11}}_{\tilde\bdA} \tilde \bdx_1+\bigg[\underbrace{\tilde \bdQ\tilde
\bdH_{12} \tilde \bdP_a}_{\tilde\bdB} ,\underbrace{\tilde \bdQ\tilde
\bdH_{12} \tilde \bdP_b}_{\tilde\bdC} \bigg] \tilde \bdx_2 +\tilde
\bdQ\tilde \bdz_1.
\end{align}
To achieve the degrees of freedom pair shown in \eqref{eq.dofpair.kleqn1} for
both users, it is sufficient to design our $\tilde \bdP$ and $\tilde \bdQ$ to
satisfy the following conditions simultaneously
\begin{enumerate}
\item $\rank(\tilde\bdA)=\M1K$,
\item $\rank([\tilde \bdP_a,\tilde \bdP_b])=M_2(K-M_1)+M_1(N_1-K)$,
\item $\tilde \bdB=\bdzeros$,
\item $\tilde \bdC=\bdzeros$.
\end{enumerate}
We propose to use the following realizations:
\begin{align}
 \bdQ^\spF&= [\vand_{K}(1, \omega_{K},\dots,\omega^{M_{1}-1}_{K})]^T,
 	\label{eq.f.Q}\\
 \bdP_a^\spF&=\vand_{K}(\omega_{K}^{-M_1},\omega_{K}^{-(M_1+1)},\dots,
 	\omega^{-(K-1)}_{K}).  \label{eq.f.Pa}\\
 \bdP_b^\spF&=(\bdQ^\spF)^\dg, \label{eq.f.Pb}\\
  \bdP_a^\spS&=\bdI_{M_2}, \label{eq.s.Pa}
\end{align}
\begin{align}
 \bdP_b^\spS&=[\bdI_{N_1-K};\bdzeros], \label{eq.s.Pb}\\
  \bdQ^\spS&=\text{null}(\bdH_{12}\bdP_b^\spS)^T, \label{eq.s.Q}
\end{align}
where \eqref{eq.s.Q} means that $\bdQ^\spS \bdH_{12}\bdP_b^\spS = \bdzeros$.
Here, we choose $((\bdQ^\spF)^\dg,\bdP_a^\spF)$ to be a size $K\times K$ IFFT
matrix, which offers the same frequency domain explanation as discussed in
\secref{sec.dis}; see also \figref{fig.spacefreq}. It is trivial to see
$\tilde \bdB=\bdzeros$. In other words, receiver 1 will simply ignore the
signal in the last $K-M_1$ frequencies and only using the signal in the first
$K$ frequencies to decode his own message. Therefore, $\tilde \bdP_a$ contains
the interference directions from all the antennas of transmitter 2 but only in
certain frequencies. Now, after applying the frequency nulling, there are
$N_1K$ dimensions remaining, which contain both user 1's message and the
message of user 2 that is transmitted by $\tilde \bdP_b$. Among all the $N_1K$
dimensions, receiver 1 only requires $M_1K$ dimensions to decode his own
message, while leaving additional $K(N_1-M_1)$ dimensions for user 2. Here we
choose one possible way of decomposing the remaining dimensions. Transmitter 2
sends some messages in the first $M_1$ frequencies but only though $N_1-K$
antennas, as shown in \eqref{eq.s.Pb}. Notice that
\begin{align}
\tilde\bdC&=\tilde \bdQ\tilde
\bdH_{12} \tilde \bdP_b\nonumber\\
&=(\bdQ^\spF\otimes \bdQ^\spS)(\bdI_K\otimes \bdH_{12})(\bdP_b^\spF
	\otimes \bdP^\spS_b)\\
&=(\bdQ^\spF\bdP_b^\spF)\otimes(\bdQ^\spS\bdH_{12}\bdP^\spS_b)
\end{align}
which means that the choice of $\bdQ^\spS$ as given in \eqref{eq.s.Q} is
sufficient to set $\tilde\bdC=0$. It is clear that for the interference signal
sent via $\tilde \bdP_b$, receiver 1 only need to do spatial zero-forcing in
our scheme, which can be seen from the fact $\bdQ^\spF\bdP_b^\spF=\bdI_K$ due
to \eqref{eq.f.Pb}.

To satisfy the second condition, notice that $\rank(\tilde \bdP_a)=M_2(K-\M1)$
and $\rank(\tilde \bdP_a)=M_{1}(N_1-K)$, it is sufficient to show that $\tilde
\bdP_a \perp \tilde \bdP_b$, which is obvious as
\begin{align}
\tilde\bdP_b^\dg\tilde\bdP_a= (\bdQ^\spF \bdP_a^\spF)
\otimes((\bdP^\spS_b)^\dg\bdI_{\M2})=\bdzeros
\end{align}
because $\bdQ^\spF \bdP_a^\spF=\bdzeros$. This is not surprising as the signal
of user 2 transmitted via $\tilde\bdP_a$ and $\tilde\bdP_b$ are orthogonal in
frequency domain. The remaining part is to show the first condition holds,
which is true because here $\tilde\bdA$ has the same structure as
$\tilde\bdA'$ of \eqref{eq.eqch.u1} with $N_1$ replaced by $K$ and $\bdh_i$
replaced by $\bdQ^\spS_{K\times N_1}\bdh_i$.

\subsection{Discussion}

\begin{figure*}
\centering
\includegraphics[width=0.9\textwidth]{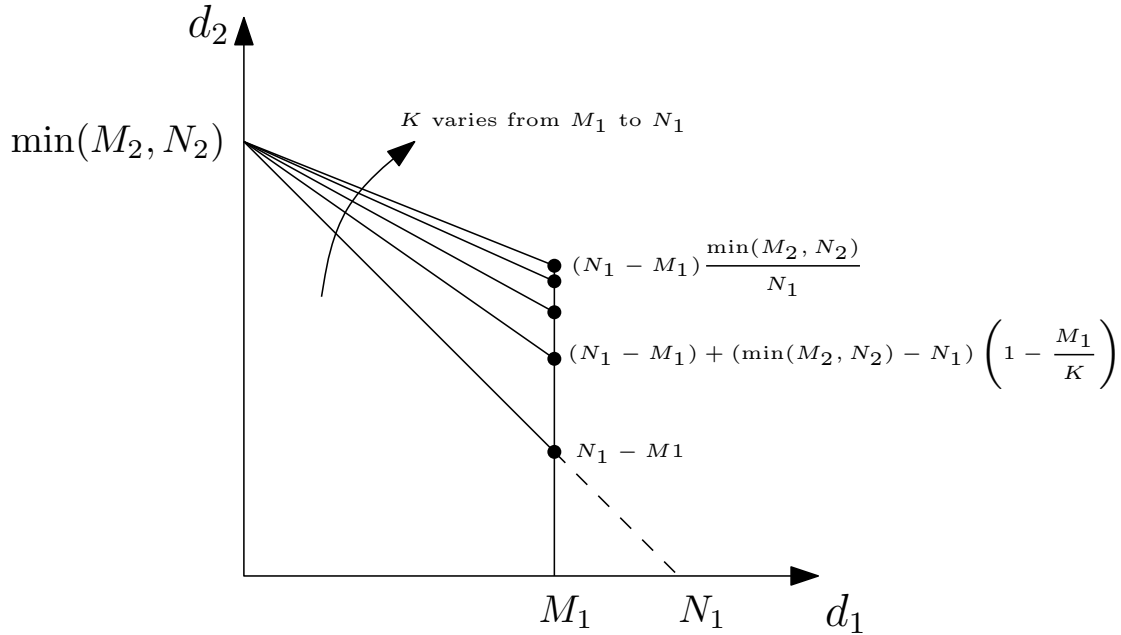}
\caption{The benefit of antenna mode switching on the DoF region, in the case
of $M_1<N_1<\min(M_2,N_2)$.}
\label{fig.gain}
\end{figure*}

It is not surprising that when $K=N_1$, \eqref{eq.modes.outer.1} implies
\begin{align}
d_1+\frac{N_1}{\min(M_2,N_2)}d_2\leq N_1
\end{align}
which is the same as \eqref{eq.nocsit.upper.m} and that in \cite[Theorem
3]{hjsv09} when $M_1< N_1 < \min(M_2,N_2)$. For the scheme that we discussed
above, $\tilde \bdP_b$ disappears and it is the DoF achievable scheme that we
developed in \secref{sec.needmode}. In addition, when $K=M_1$,
\eqref{eq.modes.outer.1} becomes \eqref{eq.dofzic.simple} and $\tilde \bdP_a$
disappears, the general scheme reduces to the DoF-optimal spatial zero-forcing
as shown in \cite{yzdg10a}. Hence, for one extra mode at transmitter 1, we can
further align $\min(M_2,N_2)-N_1$ streams of interference over $K$ slots. The
incremental gain per slot is reduced when $K$ increases; see
\figref{fig.gain}. Our result reveals the fundamental benefit that can be
obtained from reconfigurable antenna modes when there is no CSIT and $M_1< N_1
< \min(M_2,N_2)$. In addition, combining with the known results, we know that
in order to achieve the DoF region of two-user FIC and ZIC, zero-forcing in
frequency and spatial domains suffice regardless of the CSIT assumption.

\section{Conclusions}\label{sec.conc}

We derived the exact DoF region for the MIMO Z and full interference channels
when perfect channel state information is available at receivers, including i)
the Z interference channel with channel state information at the transmitter;
ii) the Z and full interference channel without channel state information at
the transmitter, but with reconfigurable antennas at the transmitters. For
both FIC and ZIC, when the number of antenna modes $K$ at the transmitter with
the reconfigurable antennas is not less than the number of receive antennas at
the corresponding receiver, the DoF region is maximized and no longer depends
on the number of antenna modes. Otherwise, each additional antenna mode can
bring extra gain in the DoF region when $\M1 < \N1 < \min(\M2,\N2)$ for both
FIC and ZIC, and when $\M2 < \N2 < \min(\M1,\N1)$ for FIC. The incremental
gain diminishes as $K$ increases.

The achievability schemes we designed for the reconfigurable antenna cases
rely on time expansion and joint beamforming and nulling over the
time-expanded channel. Interestingly, they also bear a space-frequency coding
interpretation. We completely characterized the DoF regions for both Z and
full interference channels when transmitter antenna mode switching is allowed.
Our result can specialize to previously known cases when there is no antenna
mode switching by simply setting the number of antenna modes equal to the
number of transmit antennas. Our work reveals how the channel variation
introduced by the extra antenna mode switching brings benefits in the sense of
the DoF region.

\section*{Acknowledgment} The authors would like to thank one anonymous
reviewer of an early draft of our work for suggesting that we consider antenna
mode associated with each antenna.

\linespread{1.32}\normalsize
\bibliographystyle{IEEEtran}
\bibliography{refs}

\begin{thebibliography}{10}
\providecommand{\url}[1]{#1}
\csname url@rmstyle\endcsname
\providecommand{\newblock}{\relax}
\providecommand{\bibinfo}[2]{#2}
\providecommand\BIBentrySTDinterwordspacing{\spaceskip=0pt\relax}
\providecommand\BIBentryALTinterwordstretchfactor{4}
\providecommand\BIBentryALTinterwordspacing{\spaceskip=\fontdimen2\font plus
\BIBentryALTinterwordstretchfactor\fontdimen3\font minus
  \fontdimen4\font\relax}
\providecommand\BIBforeignlanguage[2]{{%
\expandafter\ifx\csname l@#1\endcsname\relax
\typeout{** WARNING: IEEEtran.bst: No hyphenation pattern has been}%
\typeout{** loaded for the language `#1'. Using the pattern for}%
\typeout{** the default language instead.}%
\else
\language=\csname l@#1\endcsname
\fi
#2}}

\bibitem{carl75}
A.~Carleial, ``A case where interference does not reduce capacity
  {(Corresp.)},'' \emph{{IEEE} Trans. Inform. Theory}, vol.~21, no.~5, pp.
  569--570, May 1975.

\bibitem{sato81}
H.~Sato, ``The capacity of the {Gaussian} interference channel under strong
  interference {(Corresp.)},'' \emph{{IEEE} Trans. Inform. Theory}, vol.~27,
  no.~6, pp. 786--788, June 1981.

\bibitem{gaco82}
A.~El~Gamal and M.~Costa, ``The capacity region of a class of deterministic
  interference channels {(Corresp.)},'' \emph{{IEEE} Trans. Inform. Theory},
  vol.~28, no.~2, pp. 343--346, Feb. 1982.

\bibitem{ettw08}
R.~H. Etkin, D.~N.~C. Tse, and H.~Wang, ``Gaussian {interference} {channel}
  {capacity} to {within} {one} {bit},'' \emph{{IEEE} Trans. Inform. Theory},
  vol.~54, no.~12, pp. 5534--5562, Dec. 2008.

\bibitem{shkc08c}
X.~Shang, G.~Kramer, and B.~Chen, ``Outer bound and noisy-interference sum-rate
  capacity for symmetric {Gaussian} interference channels,'' in
  \emph{Information Sciences and Systems, 2008. CISS 2008. 42nd Annual
  Conference on}, 2008, pp. 385--389.

\bibitem{srve08c}
V.~Sreekanth~Annapureddy and V.~Veeravalli, ``{Sum capacity of the {Gaussian}
  interference channel in the low interference regime},'' in \emph{Information
  Theory and Applications Workshop, 2008}, 2008, pp. 422--427.

\bibitem{mokh08c}
A.~Motahari and A.~Khandani, ``Capacity bounds for the {Gaussian}
  {interference} {channel},'' in \emph{Proc. IEEE Intl. Symp. on Info. Theory},
  2008, pp. 250--254.

\bibitem{gbdt08}
G.~Bresler and D.~Tse, ``{The two-user Gaussian interference channel: a
  deterministic view},'' \emph{European Trans. on Telecomm.}, vol.~19, no.~4,
  pp. 333--354, Apr. 2008.

\bibitem{gbpt08}
\BIBentryALTinterwordspacing
G.~Bresler, A.~Parekh, and D.~Tse, ``{The approximate capacity of the
  many-to-one and one-to-many gaussian interference channels},'' 2008.
  [Online]. Available: \url{http://arxiv.org/abs/0809.3554v1}
\BIBentrySTDinterwordspacing

\bibitem{sadt09}
\BIBentryALTinterwordspacing
S.~Avestimehr, S.~Diggavi, and D.~Tse, ``{Wireless network information flow: a
  deterministic approach},'' 2009. [Online]. Available:
  \url{http://arxiv.org/abs/0906.5394}
\BIBentrySTDinterwordspacing

\bibitem{avvv09}
\BIBentryALTinterwordspacing
V.~S. Annapureddy and V.~V. Veeravalli, ``{Sum capacity of MIMO interference
  channels in the low interference regime},'' 2009. [Online]. Available:
  \url{http://arxiv.org/abs/0909.2074}
\BIBentrySTDinterwordspacing

\bibitem{sckp09}
\BIBentryALTinterwordspacing
X.~Shang, B.~Chen, G.~Kramer, and H.~V. Poor, ``{Capacity regions and sum-rate
  capacities of vector gaussian interference channels},'' 2009. [Online].
  Available: \url{http://arxiv.org/abs/0907.0472}
\BIBentrySTDinterwordspacing

\bibitem{gjjv03}
A.~Goldsmith, S.~Jafar, N.~Jindal, and S.~Vishwanath, ``Capacity limits of
  {MIMO} channels,'' \emph{IEEE Journal on Selected Areas in Communications},
  vol.~21, no.~5, pp. 684--702, May 2003.

\bibitem{caja08}
V.~Cadambe and S.~Jafar, ``Interference {alignment} and {degrees} of {freedom}
  of the {K} user {interference} {channel},'' \emph{{IEEE} Trans. Inform.
  Theory}, vol.~54, no.~8, pp. 3425--3441, Aug. 2008.

\bibitem{jafa07}
S.~Jafar and M.~Fakhereddin, ``Degrees of {freedom} for the {MIMO}
  {interference} {channel},'' \emph{{IEEE} Trans. Inform. Theory}, vol.~53,
  no.~7, pp. 2637--2642, July 2007.

\bibitem{jash08}
S.~Jafar and S.~Shamai, ``Degrees of {freedom} {region} of the {MIMO} {X}
  {channel},'' \emph{{IEEE} Trans. Inform. Theory}, vol.~54, no.~1, pp.
  151--170, Jan. 2008.

\bibitem{hjsv09}
\BIBentryALTinterwordspacing
C.~Huang, S.~Jafar, S.~Shamai, and S.~Vishwanath, ``On {degrees} of {freedom}
  {region} of {MIMO} {networks} without {CSIT},'' 2009. [Online]. Available:
  \url{http://arxiv.org/abs/0909.4017}
\BIBentrySTDinterwordspacing

\bibitem{zhgu09c}
Y.~Zhu and D.~Guo, ``Isotropic {MIMO} interference channels without {CSIT:}
  {The} loss of degrees of freedom,'' in \emph{Communication, Control, and
  Computing, 2009. Allerton 2009. 47th Annual Allerton Conference on}, 2009,
  pp. 1338--1344.

\bibitem{cvmv09}
\BIBentryALTinterwordspacing
C.~S. Vaze and M.~K. Varanasi, ``{The degrees of freedom regions of MIMO
  broadcast, interference, and cognitive radio channels with no CSIT },'' 2009.
  [Online]. Available: \url{http://arxiv.org/abs/0909.5424}
\BIBentrySTDinterwordspacing

\bibitem{sckp09c}
\BIBentryALTinterwordspacing
X.~Shang, B.~Chen, G.~Kramer, and H.~V. Poor, ``{MIMO Z-interference channels:
  capacity under strong and noisy interference},'' 2009. [Online]. Available:
  \url{http://arxiv.org/abs/0911.4530}
\BIBentrySTDinterwordspacing

\bibitem{yzdg09}
\BIBentryALTinterwordspacing
Y.~Zhu and D.~Guo, ``{Ergodic fading one-sided interference channels without
  state information at transmitters},'' 2009. [Online]. Available:
  \url{http://arxiv.org/abs/0911.1082}
\BIBentrySTDinterwordspacing

\bibitem{jafar09}
\BIBentryALTinterwordspacing
S.~A. Jafar, ``{Exploiting channel correlations - simple interference alignment
  schemes with no CSIT},'' 2009. [Online]. Available:
  \url{http://arxiv.org/abs/0910.0555}
\BIBentrySTDinterwordspacing

\bibitem{cwtgs10}
\BIBentryALTinterwordspacing
C.~Wang, T.~Gou, and S.~A. Jafar, ``{Aiming perfectly in the dark - blind
  interference alignment through staggered antenna switching},'' 2010.
  [Online]. Available: \url{http://arxiv.org/abs/1002.2720}
\BIBentrySTDinterwordspacing

\bibitem{yzdg10a}
\BIBentryALTinterwordspacing
Y.~Zhu and D.~Guo, ``{The Degrees of Freedom of MIMO Interference Channels
  without State Information at Transmitters},'' 2010. [Online]. Available:
  \url{http://arxiv.org/abs/1008.5196}
\BIBentrySTDinterwordspacing

\bibitem{jist01}
T.~Jiang, N.~Sidiropoulos, and J.~{ten Berge}, ``Almost-sure identifiability of
  multidimensional harmonic retrieval,'' \emph{Signal Processing, IEEE
  Transactions on}, vol.~49, no.~9, pp. 1849--1859, Sept. 2001.

\end{thebibliography}

\end{document}